\documentclass[12pt]{amsart}
\usepackage{epsfig,amssymb,amsfonts}
\usepackage[latin1]{inputenc} 
\def\eps{\epsilon}

\def\bek{\beta^{-k/2}}

\def\DLie{\nabla^{a}_{x\mu}}
\def\Tr{\mbox{Tr}\;}
\def\Real#1{\mbox{Re\,}({#1})}

\def\Dag#1{{#1}^\dagger}

\newcommand{\der}[2]{\dfrac{d #1}{d #2}}

\def\half{\mbox{\small $\frac{1}{2}$}}

\def\non{\nonumber}
\def\beq{\begin{equation}}
\def\eeq{\end{equation}}
\def\bit{\begin{itemize}}
\def\eit{\end{itemize}}
\def\beqn{\begin{eqnarray}}
\def\eeqn{\end{eqnarray}}
\newcommand{\ba}         {\begin{eqnarray}}
\newcommand{\ea}         {\end  {eqnarray}}
\newcommand{\ban}        {\begin{eqnarray*}}
\newcommand{\ean}        {\end  {eqnarray*}}

\begin{document}

\begin{titlepage}
\begin{flushright}
     UPRF-2004-13\\
     HU-EP-04/55 \\
     SFB/CPP-04-53 \\
     October 2004
\end{flushright}
\par \vskip 10mm
\begin{center}
{\Large \bf Numerical Stochastic Perturbation Theory \\
for full QCD}
\end{center}
\par \vskip 2mm
\begin{center}
{\bf F.\ Di Renzo}$\,^a$
and {\bf L.\ Scorzato}$\,^{b}$\\[.5 em]
$^a\,${\sl Dipartimento di Fisica, Universit\`a di Parma} \\
and {\sl INFN, Gruppo Collegato di Parma, Parma, Italy}\\[.5 em]
$^b\,${\sl Institut f\"ur Physik, Humboldt Universit\"at, Berlin, Germany} \\[.5 em]
\vskip 2 mm
\end{center}
\par \vskip 2mm
\begin{center} { \large \bf Abstract} 
 \end{center}
\begin{quote}
{\sl 
We give a full account of the Numerical Stochastic Perturbation 
Theory method for Lattice Gauge Theories. 
Particular relevance is given to the inclusion
of dynamical fermions, which turns out to be surprisingly cheap 
in this context. We analyse the underlying 
stochastic process and discuss the convergence properties. 
We perform some benchmark calculations and - as a byproduct -
we present original results for Wilson loops and the 3-loop 
critical mass for Wilson fermions.
}
\end{quote}
\end{titlepage}

\section{Introduction}

Perturbation Theory applied to Quantum Field Theories 
in Lattice regularization is very difficult. 
One has to face the difficulty of handling complicated trigonometric 
functions defined in the Brillouin zone (the momentum space on the lattice). 
Things become even more cumbersome in the case of Lattice Gauge Theories  
due to appearance of new Feynman vertices at any new perturbative order. 
This makes computations extremely difficult at 2-loops level and 
virtually unfeasible beyond that.
Ideally Perturbation Theory would not be necessary on the lattice.
However, in practice, it is needed for a number of reasons:
to match perturbative results obtained
in a continuum regularization (typically ${\overline{\rm MS}}$) 
with the non-perturbative ones from the lattice; to compute perturbative 
renormalization factors of bare parameters and operators; to determine the 
so-called improvement coefficients for lattice actions and operators. 
In recent years much progress has come 
from a non--perturbative implementation of many of these tasks. Still, 
perturbative results are needed when non-perturbative results are not 
available (and difficult to achieve). Also, one would like to gain insight 
by comparing perturbative and non-perturbative results (and a fair 
comparison actually calls for going beyond one loop). In the end, one 
knows that Perturbation Theory has to be reliable in a given limit of the theory and it 
is important to carefully assess what the results are in this limit. 
For a recent review of Perturbation Theory on the lattice see 
\cite{CapitaniRev}. A pioneering high order (three loops) computation in QCD with 
Wilson fermions can be found in \cite{SandraTomeu}.

Numerical Stochastic Perturbation Theory (NSPT) 
was introduced \cite{DRMMOLatt94,DRMMO94} as a numerical application of 
Stochastic Quantization \cite{PaWu,DH} and successfully applied 
\cite{8Loop,OPEpr,MassTerm,VolFin} to perform high order perturbative 
calculations in Lattice Gauge Theories (LGT).
The basic idea is to integrate on a computer the differential equations of 
Stochastic Perturbation Theory. This results in a slight modification of
a non perturbative Langevin algorithm. 
Until very recently the applications of NSPT to LGT were limited to 
the quenched case. Given the similarity between NSPT and a non-perturbative Monte Carlo, one could fear that unquenched NSPT would cost some order of magnitude 
more than the quenched case, just like the non perturbative algorithm from which it is derived. The main purpose of this paper is to describe
the unquenched version of NSPT for LGT and present some benchmark analysis. 
The good message is that full NSPT is not much more expensive than quenched NSPT, 
provided a fair efficiency is granted for a basic tool, namely the Fast Fourier Transform. 
In fact the perturbative expansion of the inverse fermion matrix only requires 
the inverse of the {\em free} fermion matrix. We will show how to efficiently 
exploit this property.

Since the first introduction of NSPT in \cite{DRMMOLatt94,DRMMO94}, much experience
has been collected which brought to a better understanding 
of the underlying stochastic process.
The interest about the stochastic process was strongly motivated
by the observation that, when applied to very simple (0-dim) models, NSPT
led to very large (and not normally distributed) fluctuations
at high orders. These models were studied in \cite{AdROS}. Here
we consider specifically the case of LGT. In order to rely on NSPT as a 
numerical method to extract physical results, one needs to know under 
which conditions the stochastic process converges to a limit distribution 
and assess - as much as possible - the properties of the 
limit distribution and the rate of convergence. 
We will see that a limit distribution exists, if some kind of gauge fixing
has been introduced and if the fields do not contain zero modes.
The assessment of the properties of the limit distribution - which is crucial - 
was already discussed in \cite{AdROS}. There it was shown that LGT do not display
the pathological behaviour of the 0-dimensional models, and an argument
was given to understand why this happens. As a benchmark for error analysis, 
in this paper we collect results for a large set of Wilson loops and
show how the fluctuations depend on the loop size and
the perturbative order. 

It is often useful to compute gauge dependent quantities
in a fixed gauge. In the context of NSPT a form of gauge fixing 
is always imposed by a stochastic term inspired to the gauge fixing procedure 
introduced by Zwanziger in \cite{Zwa} (actually our prescription is a perturbative 
form of \cite{Rossi:hv}). However, it is not clear what is the perturbative 
relation between this gauge fixing and the more popular ones. Still, we can 
show a simple way to recover Landau gauge fixing. We can also perform computations in covariant gauges without introducing ghost fields. The Faddeev--Popov determinant is
represented by the same technique used for the fermionic one. 

Another interesting feature of NSPT is the 
possibility of computing, with the same effort, perturbative expansion around
non trivial vacua. The vacua configurations may even be known only numerically. 

In section 2 we briefly review how NSPT can be naturally derived from the context 
of Stochastic Quantization and how this can be specified to the case of LGT.
In section 3 we concentrate on the analysis of the NSPT stochastic process 
for LGT: we first explain why a stochastic
gauge fixing is needed and how it can be practically implemented; then we
show how the modes with zero momenta must be regularized in 
this context. These corrections allow us to prove
the convergence of the stochastic process.
We conclude the section with a comment on simulation times
and their dependence on the lattice volume and the maximum
perturbative order. In the same spirit of assessing the efficiency of the method 
we also perform the error analysis of a large set of Wilson loops in the quenched 
approximation, discussing the dependence on the loop size and the perturbative order.
Section 4 is devoted to the illustration of some special applications. 
First we show how to evaluate in perturbation theory gauge dependent quantities in Landau 
gauge. We also comment on the strategy for the implementation of the Faddeev--Popov 
mechanism for general covariant gauges. Some more comments on this subject are 
differed to the following discussion of fermions, since it relies on the same 
technique. We then consider perturbative expansions around non trivial vacua. 
Section 5 is devoted to the inclusion of dynamical fermions, while in section 6 
we present benchmark unquenched computations for $n_f=2$ 
Wilson fermions, \emph{i.e.} Wilson loops and the critical mass to the third order. 
In each section we also provide directions for a practical
implementation of the method in a Monte Carlo simulation.
In section 7 we give our conclusions.
\section{From Stochastic quantization to NSPT for Lattice Gauge Theories}
\subsection{Stochastic Quantization}
NSPT was introduced \cite{DRMMOLatt94,DRMMO94} as a numerical application of 
Stochastic Quantization \cite{PaWu,DH}. It is well known that from the latter one 
can formulate a Stochastic Perturbation Theory, which converges (in a 
sense that will be clear in a moment) to the standard perturbative expansion 
of field theories. NSPT is nothing but the numerical implementation of this 
program. In order to fix the main ideas and introduce the notation 
let us remind the fundamental assertion of Stochastic Quantization. 
One starts with an action for a field theory $S[\phi]$ and aims at computing 
expectation values with respect to the path integral measure, \emph{i.e.} 
\[
	\langle O[\phi] \rangle = 
	\frac{\int D\phi \;\,O[\phi]\;\, e^{-S[\phi]}}{\int D\phi \;\,e^{-S[\phi]}}.
\]
The basic fields of the theory are now given an extra degree of freedom $t$ 
($\phi(x) \mapsto \phi_\eta(x;t)$). One can think of it as a fictitious 
(stochastic) time in which an evolution takes place according to the 
Langevin equation
\begin{equation} \label{BasicLANG}
\der{\phi_\eta(x;t)}{t} = - \frac{\partial S[\phi]}{\partial \phi_\eta(x;t)} + \eta(x;t).	
\end{equation}
The first term is the \emph{drift} term given by the equations of motion (which 
would drive the field to the classical solution). One adds to the latter a 
Gaussian noise
\[
	\eta(x;t): \;\;\;\;\; \langle \eta(x,t) \;\, \eta(x',t') {\rangle}_\eta = 
	2 \, \delta(x-x') \;\, \delta(t-t')
\]
where
\[
\langle \ldots {\rangle}_\eta = \frac{\int D\eta(z,\tau) \, \ldots \, e^{- \frac{1}{4} \int dz d\tau \eta^2(z,\tau)} }{\int D\eta(z,\tau) \, e^{- \frac{1}{4} \int dz d\tau \eta^2(z,\tau)}}.
\]
$\eta$ is the source of the stochastic nature of the process (this motivates 
the notation $\phi_\eta$). 
The natural expectation values to compute in this framework are with respect 
to this noise. The main 
assertion of Stochastic Quantization is that \cite{PaWu} 
\begin{equation} \label{basicSQ}
	\langle O[\phi_\eta(x_1;t) \ldots \phi_\eta(x_n;t)] {\rangle}_\eta
	{\rightarrow}_{t \rightarrow \infty} 
	\langle O[\phi(x_1) \ldots \phi(x_n)] \rangle,
\end{equation}
which means that \emph{in the limit of the stochastic time going to infinity expectation 
values of the field with respect to the Gaussian noise converge to the 
functional integral expectation values} one is interested in\footnote{Notice that 
all the fields are evaluated at the same stochastic time in Eq.~(\ref{basicSQ}).}. 
One can now move to Stochastic Perturbation Theory (SPT) by noting that 
$S[\phi]$ can be written as a sum $S[\phi] = S_0[\phi] + S_I[\phi,g]$, in which 
an interaction part $S_I$ (which is a function of a coupling constant $g$) has been 
singled out. For the free field action $S_0[\phi]$ an integral solution for 
the Langevin equation exists, which is written (often in momentum space) 
in terms of a Green function. For the full theory, the Langevin equation can be converted into an integral equation, whose iterative solution as a series in the coupling constant actually results in SPT. This is the original approach to SPT, for which the interested reader can refer to \cite{PaWu}. Notice that this same approach can be recovered by looking for a solution of the Langevin equation as a (formal) series 
in the coupling constant
\begin{equation}
	\phi_\eta(x;t) = \phi_\eta^{(0)}(x;t) + \sum_{n>0} g^n \phi_\eta^{(n)}(x;t).
\end{equation}
If one plugs this expansion in the Langevin equation, the latter gets translated 
into a hierarchy of equations, which one can exactly truncate at any given order. Their solutions
are the same integral expressions mentioned above.
In the simple case of a bosonic 
non-gauge theory\footnote{Notice that diagrammatic arguments are not conclusive in the 
case of gauge theories.} it is possible to show that stochastic diagrams are 
generated, which can be evaluated in the $t \rightarrow \infty$ limit. In this limit they 
reconstruct the contribution of the standard Feynman diagrams \cite{GrHuff}. 
Another, more formal, proof of the equivalence of SPT and 
standard field theoretic Perturbation Theory relies on the formalism of the 
Fokker-Planck equation\footnote{Within this framework the gauge theories case is 
neatly treated once the Stochastic Gauge Fixing term has been added.}: 
we will sketch very briefly what is the 
content of \cite{Flor}, to which we refer the interested reader. 
Let us first of all remind that the fundamental assertion contained in 
Eq.~(\ref{basicSQ}) can be rephrased in 
terms of a distribution function of the field. The first step is the introduction 
of a stochastic time dependent distribution function for the field according to 
\begin{equation}
	\langle O[\phi_\eta(t)] {\rangle}_\eta = 
	\frac{\int D\eta \, O[\phi_\eta(t)] \, e^{- \frac{1}{4} \int dz d\tau \eta^2(z,\tau)} }{\int D\eta \, e^{- \frac{1}{4} \int dz d\tau \eta^2(z,\tau)}} = 
	\int D\phi \; O[\phi] \, P[\phi,t].
\end{equation}
The Langevin equation Eq.~(\ref{BasicLANG}) can then be traded for the so-called 
Fokker--Plank equation which expresses the (stochastic) time derivative of $P[\phi,t]$ 
\begin{equation}
	\dot{P}[\phi,t] = \int dx \; \frac{\delta}{\delta\phi(x)} \left( \frac{\delta S[\phi]}{\delta\phi(x)} + \frac{\delta}{\delta\phi(x)} \right) P[\phi,t]. 
\end{equation}
The essential steps taken in \cite{Flor} can be summarized as in the following.\\

\begin{itemize}
\item 
$P[\phi,t]$ is expanded as a power series in the coupling of the theory
\[
 P[\phi,t] = \sum_{k=0} g^k P_k[\phi,t]. 
\] 
This also turns the Fokker--Plank equation into a hierarchy 
of equations. By inspecting these, the following results can be obtained. \\

\item
One can prove that 
$P_0[\phi,t] {\rightarrow}_{t \rightarrow \infty} P_0^{eq}[\phi] = 
\frac{e^{-S_0[\phi]}}{Z_0}$, \emph{i.e.} $P_0[\phi,t]$ converges to the free field 
path integral measure. \\

\item
In a convenient weak sense it is also true that 
$P_k[\phi,t] {\rightarrow}_{t \rightarrow \infty} P_k^{eq}[\phi]$. \\

\item
The various $P_k^{eq}[\phi]$ are related by a set of relations in which one 
recognizes the Schwinger--Dyson equations. Since the solutions of the 
Schwinger--Dyson equations 
are unique in Perturbation Theory, one has recovered the standard field theoretic 
perturbative expansion. \\

\item
In the case of gauge theories, a key role in getting through the steps sketched 
above is played by the so called Stochastic Gauge Fixing. \\
\end{itemize}

This ends our introduction to Stochastic Quantization and Stochastic Perturbation 
Theory. As already said, NSPT is a numerical implementation of SPT, which 
in particular we apply to LGT. The numerical nature opens the door to all the 
problems of numerical stability. It also 
motivates a peculiar treatment of the question of convergence properties of NSPT 
stochastic process for LGT, which we will address in section 3. 
As motivated in the introduction, this issue 
is a practically important one. This will also lead us to the introduction of 
Stochastic Gauge Fixing, which, as said, is crucial for the results of 
\cite{Flor}. We now show the practical implementation of 
NSPT for LGT. \\

\subsection{NSPT for Lattice Gauge Theories}

Consider the Euclidean Wilson action (for the gauge group $SU(N_c)$):
\[
S_G \, = \, - \frac{\beta}{2 N_c} \, \sum_P \, \mbox{Tr} \left( U_P + U^{\dag}_P 
\right).
\]
Just as before, Stochastic Quantization amounts to considering a set of fields 
$U_{x\mu}(t;\eta)$ which, besides the usual dependence on space-time ($x\in {\mathbb R}^4$) 
and direction ($\mu=0\ldots3$), also depend on the
{\em stochastic time} $t\in {\mathbb R}$ and (parametrically) on a random field $\eta$. 
The Langevin equation for the $U$ fields now reads 
\begin{equation} \label{Langevin}
\frac{\partial}{\partial t} U_{x\mu}(t;\eta) \, = \, 
\left( -i \nabla_{x\mu} S_G[U] -i \eta_{x\mu}(t) \right) U_{x\mu}(t;\eta),
\end{equation}
where $\nabla_{x\mu} =  T^a \nabla^a_{x\mu} = T^a \nabla^a_{U_{x\mu}}$ is a left 
derivative on the group. We recall the definition of the Lie derivative ($V$ is an 
element of the relevant Lie group) 
\[
\nabla^a_V f(V) = \lim_{\alpha\rightarrow 0} \frac{1}{\alpha}
(f\left( e^{i \alpha T^a} V \right) \, - \, f(V) ),
\]
whose relevant property is that integration by part is admitted. The 
$T^a$ are the hermitian generators of the algebra 
($[T^a,T^b]=i f^{a b c}T^c$ and 
$\mbox{Tr}(T^a T^b)= \frac{1}{2}\delta_{a b}$). 
The $\eta(t)=T^a \eta^a(t)$ are now random fields with a Gaussian
distribution that satisfies\footnote{
We will often omit the space-time and direction indices as well as explicit $\eta$ 
dependence whenever it is reasonable to assume that no confusion can be made.
For instance $U(t) = U_{x\mu}(t;\eta)$, and $\eta^a(t)=\eta_{x\mu}^a(t)$.} 
\[
\langle \eta^a(t) {\rangle}_\eta \, = \, 0  \qquad \quad 
\langle \eta_{x\mu}^a(t) \, \eta_{y\nu}^b(t') {\rangle}_\eta \, = \, 
2 \, \delta^{ab} \, \delta_{\nu\mu}\, \delta_{y x} \, \delta(t-t'),
\]
and higher cumulants vanish. Again, one can prove \cite{PaWu,DH} that under 
these assumptions the gauge fields distribute according to the 
measure $e^{-S[U]}$ in the limit
of large $t$. In other words (if $U(t;\eta)$ is a solution of 
(\ref{Langevin}) determined by $\eta$):
\[
\lim_{t\rightarrow\infty} \langle O[U(t;\eta)] {\rangle}_\eta \, =
\frac{1}{Z} \int DU \, e^{- S_G[U]} \, O[U].
\]

Having in mind a numerical integration on a computer, 
the Langevin time can be discretized (with step $\epsilon$ and $t=n\epsilon$).
Following \cite{Batrouni}, one can check that a solution to (\ref{Langevin}) is given 
by 
\begin{equation}\label{eq:iteration}
U_{x\mu}(n+1;\eta) = e^{- F_{x\mu}[U,\eta]} \,\, U_{x\mu}(n;\eta)
\end{equation}
where
\begin{eqnarray}\label{eq:F}
F_{x\mu}[U,\eta] \, &=& \eps \nabla_{x\mu} S_G[U] + \sqrt{\eps} \, \eta_{x\mu} \\
&=& \, \frac{\epsilon \beta}{4 N_c}
\sum_{U_P \supset U_{x\mu}} \left[ \left( U_P - U^{\dag}_P \right) - 
\frac{1}{N_c} \mbox{Tr} \left( U_P - U^{\dag}_P \right) \right] + 
\sqrt{\epsilon} \, \eta_{x\mu} \non\\
\end{eqnarray}
and the matrix degrees of freedom of the random field $\eta$ are subjected 
to\footnote{In this formula the focus is on matrix indices. $z$ and $w$ are 
multi--indices collecting the space and (stochastic) time degrees of freedom.} 
\[
\langle \eta_{i,k}(z) \,\, \eta_{l,m}(w) {\rangle}_\eta \, = \, 
\left[ \delta_{il} \, \delta_{km} - \frac{1}{N_c} \, \delta_{ik} \, \delta_{lm} 
\right] \delta_{zw}. 
\]
Eq.~(\ref{eq:iteration}) is basically an Euler scheme for Eq.~(\ref{Langevin}) 
taking care of not leaving the group manifold. Being an Euler scheme, results 
are to be extrapolated linearly as $\eps\rightarrow 0$ in order to recover 
a solution of Eq.~(\ref{Langevin}). Notice also that the drift $F_{x \mu}$ is a local 
expression: in order to update the link $U_{x \mu}$ it suffices to compute the 
plaquettes $U_P$ insisting on $U_{x \mu}$. \\

One can now proceed as stated above. Since $\nabla S_G$ obviously depends on the 
coupling, also the fields $\{U_{x\mu}(t;\eta)\}$ acquire a dependence on $\beta$ 
through the Langevin equation (\ref{Langevin}).
As a consequence we can write, at least formally,
\begin{equation} \label{sost}
U_{x\mu}(t;\eta) \rightarrow 1+ \sum_{k=1} \beta^{-k/2} U^{(k)}_{x\mu}(t;\eta).
\end{equation}
The expansion starts with the unity operator. This is actually a choice 
for the vacuum configuration around which we are going to compute perturbative 
corrections. As we will see later, this is not the only possible choice. 
Due to the presence of a $\beta$ in front of the Wilson action, if one now 
plugs the expansion (\ref{sost}) in $F$, the latter starts at order $\sqrt{\beta}$ 
for the drift term, while has a zero order contribution from the random field $\eta$. 
This makes a perturbative evaluation of Eq.~(\ref{eq:iteration}) inconsistent. 
However we can redefine the time
step $\epsilon'=\epsilon\beta$, in such a way that the first non zero
contribution to $F$ is of order $\frac{1}{\sqrt{\beta}}$ and it comes both 
from the drift and from the random noise $\eta$. 

The introduction of the expansion (\ref{sost}) inside the Langevin equation 
transforms (\ref{eq:iteration}) into a system of equations, one for each 
perturbative component of the field\footnote{We move to a lighter notation and 
write for the Euler step $U' = e^{-F} U$.}:
\begin{eqnarray} \label{Lansystem}
{U^{(1)}}' & = & U^{(1)} - F^{(1)} \\ 
{U^{(2)}}' & = & U^{(2)} - F^{(2)} + \half F^{(1)\,2} - F^{(1)} U^{(1)} \nonumber \\
{U^{(3)}}' & = & 
U^{(3)} - F^{(3)} + \half (F^{(2)} F^{(1)}+F^{(1)} F^{(2)}) - 
{\mbox{\small $\frac{1}{3!}$}} 
F^{(1)\,3} \nonumber \\
& & - (F^{(2)} - \half F^{(1)\,2})\,U^{(1)}- F^{(1)} U^{(2)} \nonumber \\
\ldots & & \nonumber
\end{eqnarray} 
Again, the system of equations above can be consistently truncated anywhere. 
In fact the equation for $U^{(k)}$ only depends on fields of equal or lower
perturbative order. Moreover the dependence on $U^{(k)}$ in the $k-$th
equation is trivial. As already pointed out, $\eta$ only 
enters $F^{(1)}$, so that the first order explicitly depends on the noise, while 
higher orders are stochastic via the dependence on lower orders. Again, having 
adhered to an Euler scheme, results have to be extrapolated linearly as $\eps\rightarrow 0$.
\subsubsection{A numerical strategy}
In the end NSPT simply amounts to integrate the equations (\ref{Lansystem}) 
on a computer, \emph{i.e.} we are now going to 
sketch the strategy for a {\em perturbative} Monte Carlo. 
If we want to perform a perturbative calculation to the order
$g^n$ we need to replicate the gauge configuration $n$ times, and
evolve the whole set according to the Langevin system (\ref{Lansystem}).
An important practical advantage of this method is that it can
be coded by introducing very few modifications to a non perturbative
Monte Carlo program for a local field theory. 
We have so far discussed the perturbative expansion
of the Langevin equation. This choice is not essential: we could have
chosen other stochastic dynamics, as long as they are based on differential
equations (see for instance \cite{CR82,DK86,Kramer}).
Notice however that there is no possible perturbative expansion for an 
accept/reject Metropolis step, hence there is no NSPT analogous 
for such a kind of updating algorithm. 
In practice the mechanism of replicating the fields in perturbative orders 
consists in introducing in the field configuration the dependence on a new index 
(the perturbative order itself). Notice that any algebraic operation must be 
expanded perturbatively: 
\begin{eqnarray}\label{pertalg}
X= A + B &\rightarrow& X^{(k)} = A^{(k)} + B^{(k)}, \nonumber\\
X= A * B &\rightarrow& X^{(k)} = \sum_{j=0}^k  A^{(j)} \times B^{(k-j)}.
\end{eqnarray}
Any measurement code for any observable\footnote{As for fermionic or
gauge fixed observables see the relative sections.} can be adapted
to the present context by following the rules above.
What we are typically interested in are the coefficients of the expansion\footnote{
We adopt the notation for a generic field theory.}
\begin{equation} \label{pertOBS}
	\langle O[\sum_k g^k \phi^{(k)}_{\eta}(t)]\rangle_{\eta} = \sum_k g^k O_k(t),
\end{equation}
which also appear naturally as variables having a perturbative index and 
obeying the same multiplication rules (\ref{pertalg}).

The average over the noise fields $\eta$ is reproduced, as usual,
by averaging over a single long history, provided that one takes into
account the autocorrelation at length $n$
\[
\lim_{t\rightarrow \infty}\langle O_k(t) \rangle_{\eta} = 
\lim_{T\rightarrow \infty} 1/T \sum_{j=1}^T O_k(j n). 
\]

Finally notice that, if a non perturbative code is written by means of 
an object oriented language (where, typically, matrix operations are
realized through the definition of suitable operators acting on classes),
then the modifications just described could be really minor ones.

A drawback of the method is the high request for memory (of course also the 
Floating Point operations requested by (\ref{pertalg}) increase with the 
perturbative order). These requirements however 
are achievable given our current computing power.

The implementation of NSPT for fermions is quite different, and we will discuss 
it in section \ref{fermions}.
\subsubsection{The point of view of the algebra}
The expansion (\ref{sost}) may appear quite strange. As a matter of fact, 
one is quite familiar with a non perturbative formulation of Wilson action 
whose fields are the $U_{x\mu}$, which take their values in the group. On 
the other side, Perturbation Theory amounts to taking into accounts 
fluctuations around a vacuum configuration, which naturally results in 
considering the Lie algebraic fields $A_{x\mu}$. NSPT does not perform 
anything strange with this respect. One can also work with field variables 
which live in the algebra, where our perturbative expansion reads\footnote{One 
could observe that a better notation for the $A$ field would be 
$A_{x\mu}(t;\eta) \rightarrow \sum_{k=1} \beta^{-k/2} A^{(k-1)}_{x\mu}(t;\eta)$. 
This would enlighten the fact the first term in the expansion of $A$ is actually a 
free field. Still, the bookkeeping of indices is more convenient in our notation, 
which is the reason for our choice.} 
\begin{equation} \label{pertfields}
A_{x\mu}(t;\eta) \rightarrow \sum_{k=1} \beta^{-k/2} A^{(k)}_{x\mu}(t;\eta).
\end{equation}
This is perfectly equivalent. There is of course a perturbative relation
between the two expansions: 
{\footnotesize
\begin{eqnarray}\label{eq:AU}
A & = & \log (U ) =  \log \left( 1 + \sum_{k>0}
\beta^{-\frac{k}{2}} U^{(k)}
\right) \\
& = & \frac{1}{\sqrt{\beta}} U^{(1)} + \frac{1}{\beta} \left(
U^{(2)} - \frac{1}{2} U^{(1) \, 2} \right) + \left (
{\frac{1}{\beta}} \right )^{\frac{3}{2}} \left( U^{(3)} -
\frac{1}{2} \left( U^{(1)} U^{(2)} + U^{(2)} U^{(1)} \right) 
+ \frac{1}{3} U^{(1) \, 3} \right) \nonumber \\ & & + \ldots
\nonumber \\ & = & \frac{1}{\sqrt{\beta}} A^{(1)} +
\frac{1}{\beta} A^{(2)} +
\left ( {\frac{1}{\beta}} \right )^{\frac{3}{2}} A^{(3)}  
+ \ldots \nonumber
\end{eqnarray}
}
The transformation (\ref{eq:AU}) between the $U$'s and $A$'s 
variables can of course be performed exactly at any finite order $g^n$. 
We should stress once again the choice of the vacuum (background) 
configuration which is embedded in (\ref{eq:AU}). The Lie algebraic solution 
of the equations of motion is the usual (trivial) perturbative vacuum 
$A_{x\mu}=0$, so that the first non zero order is given by fluctuations 
of order $O(g)$ (\emph{i.e.}, in our preferred notation, 
$O(\frac{1}{\sqrt{\beta}}))$. This, in turn, results 
in (\ref{sost}) being expressed as $1$ plus fluctuations of order 
$O(\frac{1}{\sqrt{\beta}})$. It is worth to stress that in going from one 
notation to the other one has to handle the trivial 
series expansions of $\log(1+x)$ and $\exp(x)$, which can again be coded 
in an \emph{order by order} notation according to the rules (\ref{pertalg}). 
The constraint of unitarity on the original (not 
expanded) $U_{x\mu}$ fields is translated into the usual (anti)hermitian, 
traceless nature of the $A$'s fields 
\[
A^{(k) \, {\dag}} \, = \, - A^{(k)} \qquad
\mbox{Tr} A^{(k)} \, = \, 0 \qquad
\forall k,
\]
\emph{i.e.} the $A^{(k)}$ are still Lie algebraic fields. Notice that on the 
other hand there is no simple algebraic characterization of the $U^{(k)}$ 
fields, which are with this respect in a sense less fundamental. However, 
the transformation (\ref{eq:AU}) suffices to ensure that the fundamental relation 
$U U^\dag = 1 = U^\dag U$ is satisfied at every finite order in the coupling constant 
(via highly non linear relations). Once expansions have been made, no matter what the choice of notation is, one has effectively decompactified the formulation of the theory (but this is not of course a feature of NSPT, but of Perturbation Theory itself).

During the evolution it is necessary to periodically enforce 
the $SU(3)$ group constraints, which may be spoiled by round-off errors.
In our case this has to be done on the perturbative expansion of the gauge variables. 
The constraints are easier to enforce on the Lie algebra fields $A^{(k)}$. 
This is not the only advantage of working with variables living in the algebra.
One can also get a non irrelevant saving of memory. In fact the highest
order of the field typically only appears in observables linearly under 
trace, so that $A^{(k_{max})}$ can be omitted completely. 
On the other side the point of view of the algebra is less natural (in the 
end, Wilson action is formulated in terms of the $U$'s). 
For instance the Langevin system of equations (\ref{Lansystem}) 
gets much more involved:
\begin{eqnarray} \label{Lansystalg}
{A^{(1)}}' & = & A^{(1)} - F^{(1)} \\ {A^{(2)}}' & = &
A^{(2)} - F^{(2)} - \frac{1}{2} \left[ F^{(1)} ,
A^{(1)} \right] \nonumber \\ {A^{(3)}}' & = & A^{(3)} -
F^{(3)} - \frac{1}{2} \left[ F^{(1)} , A^{(2)} \right] -
\frac{1}{2} \left[ F^{(2)} , A^{(1)} \right] \nonumber \\ & &
+ \frac{1}{12} \left[ F^{(1)} , \left[ F^{(1)} , A^{(1)}
\right]
\right] 
+ \frac{1}{12} \left[ A^{(1)} , \left[ F^{(1)} , A^{(1)}
\right]
\right] \nonumber \\
\ldots & & \nonumber
\end{eqnarray} 
For this reason we usually prefer the expansion entailed in Eq.~(\ref{sost}), even if 
for certain operations it is useful to switch to the algebra. 
\section{Analysis of the stochastic process}
The discussion above shows that the perturbative coefficients $O_k$ in 
Eq.~(\ref{pertOBS}) correspond to well defined combinations of correlation 
functions of the stochastic processes $\phi^{(j)}(t)$. We now want to stress 
once again that 
NSPT is a \emph{numerical implementation} of Stochastic Perturbation Theory. With this 
respect not only the existence of limit distributions matters, but also the 
properties of convergence. In principle one would like to characterize as best as 
possible the stochastic processes in order to rely on NSPT as a computational tool. 
We will now restrict our attention to the case of gauge theories and 
in order to gain insight we will address in the following the study of the generic 
correlation function of $M$ perturbative components of the fields\footnote{In the following 
we will alternate the notations $A_{x\mu}$ and $A_\mu(x)$. The latter is easier to read, 
in particular when there are other indices around (as in $x_j$) and it avoids confusion 
when Fourier transformations are involved.}
\begin{equation} \label{funzcorr}
\langle\prod_{j=1}^{M} A^{(p_j)}_{\mu_j}(x_j;t) \rangle.
\end{equation}
One can think of this expression as entering the perturbative expansion of a 
generic observable. Notice however that taken by itself the previous expression is 
by far more general. After showing that all such correlations have a finite limit for
$t\rightarrow\infty$, we will collect some results about the 
rate of convergence of these expressions. As a matter of fact, this 
will give us the chance to motivate some prescriptions in the implementation 
of NSPT for LGT, in particular the Stochastic Gauge 
Fixing, which is crucial, as already said, to demonstrate the results of \cite{Flor}. 
The same questions were studied in \cite{AdROS} for some simple models. 
The purpose of this section is to analyse the peculiarities of LGT.
This will lead us to the introduction of Stochastic Gauge Fixing and
the regularization of zero modes. \\

Let us start considering the process defined 
by (\ref{Lansystem}) (or equivalently (\ref{Lansystalg})), 
with $F$ given in (\ref{eq:F}). 
It is not difficult to see that a limit distribution cannot exist.
To illustrate this point we consider the
system in a continuum regularization. The Langevin equation then 
reads ($D^{ab}_\nu$ is the gauge covariant derivative; no expansion has yet 
been made)
\[
\frac{\partial}{\partial t}A^a_{\mu}(\eta,x;t) \, = \, D^{ab}_\nu
F^b_{\nu\mu}(\eta,x;t) + \eta^a_\mu(x;t).
\]
The formal solution for the perturbative components of the fields (the 
solutions of the analogous to (\ref{Lansystem}) and (\ref{Lansystalg}))
reads (in Fourier space)
\begin{equation} \label{Sol}
A^{(n)a}_\mu(k;t) \,  = \,  T_{\mu\nu}^{ab} \int_0^t ds  \, e^{-  k^2 (t-s)}
f^{(n)b}_\nu(k,s) + L_{\mu\nu}^{ab} \int_0^t ds \, f^{(n)b}_\nu(k,s),
\end{equation}
where $T_{\mu\nu}^{ab}$ and $L_{\mu\nu}^{ab}$ 
are the abelian transverse and longitudinal projectors
\ban
T_{\mu\nu}^{ab} \, = \, 
(\delta_{\mu\nu} - \frac{k_\mu k_\nu}{k^2})\delta_{ab}, &&
\qquad \quad T_{\mu\nu}^{ab} T_{\nu\rho}^{bc} = T_{\mu\rho}^{ac}, \\
L_{\mu\nu}^{ab} \, = 
\frac{k_\mu k_\nu}{k^2}\delta_{ab}, &&
\qquad \quad L_{\mu\nu}^{ab} L_{\nu\rho}^{bc} = L_{\mu\rho}^{ac}, 
\qquad \quad T_{\mu\nu}^{ab} L_{\nu\rho}^{bc} = 0.
\ean
The function $f^{(n)}$ represents the interaction term, which only 
contains perturbative components of the field of order strictly lower than
$n$
\ban
f^{(n)a}_\nu(k;t)   &=&   g  I^{(3)   (n-1)a}_\mu(k;t)   +  g^2   I^{(4)
(n-2)a}_\mu(k;t), \\ f^{(0)a}_\nu(k;t) &=& \eta_\nu(k;t)^a. \\
\ean
Here $I^{(3) (n)a}_\mu$ and $I^{(4) (n)a}_\mu$ are the $n-$th perturbative
components of three and four gluons interaction. For instance
\[
g I_\mu^{(3)a}(k;t) \, = \, \frac{i g f^{abc}}{2 (2 \pi)^n} 
\int dp dq \, \delta(k+p+q) \, A_\nu^b(-p;t) \, A_\sigma^c(-q;t) 
\, v_{\mu\nu\sigma}^{(3)}(k,p,q),
\]
\noindent
where
\[
v_{\mu\nu\sigma}^{(3)}(k,p,q) \, = \, \delta_{\mu\nu} (k-p)_{\sigma} + 
\mbox{cyclic permutations}.
\]

A first divergence appears projecting the formal solution
(\ref{Sol}) by the longitudinal abelian projector $L_{\mu\nu}$.
Along such directions the expanded Langevin system (\ref{Lansystalg}) 
presents no damping factor $e^{-k^2 t}$ , for any of the perturbative components $A^{(n)}$ 
\[
L_{\mu\nu}^{ab}  A^{(n)b}_\mu(k;t) \,  = \,  
L_{\mu\nu}^{ab} \int_0^t ds \, f^{(n)b}_\nu(k,s). 
\]
The field $A^{(0)}$ clearly diverges like a random walk (along such degrees of freedom). 
The behaviour of the higher perturbative
components is difficult to predict in general. However, since the longitudinal
components of $A^{(n)}$ only depend on the $\{A^{(m)}|\,m<n\}$, it is natural to expect 
diverging fluctuations at any order as was indeed observed in \cite{DRMMOLatt94}.

In the lattice regularization the interaction terms $f^{(n)}$ become 
much more complicated, but the argument for expecting a divergence remains 
unchanged. This problem is not avoided if we choose to work with the  $U$ variables 
(as in (\ref{Lansystem})), instead of the algebra ones $A$ (as in (\ref{Lansystalg})). 
In fact, once the fields are perturbatively expanded, even 
the $U$ variables do not live in a compact set anymore.

A second source of divergence comes from the zero modes,
since also for these no damping factor is present (see (\ref{Sol})).
On a finite lattice the integration over momenta is substituted by a finite sum, 
and the degrees of freedom corresponding to $k=0$ give a finite 
contribution, enhanced in small lattices. This divergence results from a very 
similar mechanism as the one associated to the gauge degrees of freedom.

None of these difficulties is peculiar of stochastic quantization. The first
one appears whenever a perturbative expansion is introduced. In the traditional 
functional approach this implies the impossibility of defining the propagator, which 
is necessary to make perturbation theory meaningful. On a finite lattice, moreover, 
Perturbation Theory also faces the second problem: loops contributions are given by 
finite sums that in general entail singularities at $k=0$. In both cases the phenomenon
is related to the appearance of a zero eigenvalue in the action.
In the stochastic approach a zero eigenvalue leaves the corresponding 
degrees of freedom without an attractive force, and diverging 
fluctuations (random walks like) can appear. 
This is not a problem if one is interested in the analytical computation of
gauge invariant quantities. This was the spirit of the original proposal
of Parisi and Wu \cite{PaWu}. However, what we have in mind is a numerical computation. From this point of view, diverging fluctuations - even in intermediate quantities - can be a serious problem, as it is clear from the plots in \cite{DRMMOLatt94}.
\subsection{Stochastic gauge fixing}
Stochastic gauge fixing was introduced in \cite{Zwa} in order to
provide the stochastic quantization approach with a non perturbative
procedure of gauge fixing. The idea is to add a term to the Langevin
equation such that the evolution of gauge invariant quantities is not
affected. For instance in a continuum regularization we can write:
\[ 
\dot{A}^a_\mu(x;t) \, = \, - \frac{\delta S[A]}{\delta A_\mu^a(x;t)} 
- D_\mu^{ab} {\it V}^b[A,t] + \eta^a_\mu(x;t),
\]
where ${\it V}^a[A,t]$ is an arbitrary non gauge invariant functional.
The evolution of a generic functional $F[A]$ is given by
\[
\frac{\partial F[A]}{\partial t} \, = \, \int dx \, 
\frac{\delta F[A]}{\delta A_\mu^a(x;t)} 
\, \frac{\partial A_\mu^a(x;t)}{\partial t} ,
\]
which is not affected by the presence of ${\it V}^a[A,t]$
if $F[A]$ is gauge invariant, that is
\[ 
D_\mu^{ab} \frac{\delta F[A]}{\delta A_\mu^b(x)} \, = \, 0.
\]
One can also prove that the choice of Zwanziger
\begin{equation} \label{Zwachoice}
- D_\mu^{ab} {\it V}^b \, = \, \frac{1}{\alpha} 
\, D_\mu^{ab} \partial_\nu A_\nu^b
\end{equation}
introduces a force that keeps limited the norms of the gauge fields.

When we use a lattice regularization and the gauge variables live in the group, 
a generic gauge transformation has the form 
\begin{equation} \label{gaugetransf}
U'_{x\mu}  =
e^{w_{x}} \,\, U_{x\mu} \,\, e^{- w_{x+\hat{\mu}}},
\end{equation}
where $w$ is a field defined on the lattice sites and it takes values in 
the algebra. In this case a choice corresponding to (\ref{Zwachoice}) 
for the variation of the gauge field (modulo lattice artifacts) is\footnote{Here $\partial^L$ 
is the backward derivative.}
\begin{equation}\label{GT}
\left( 0 < \alpha < 1 \right) \qquad 
w_{x} = - \alpha 
\sum_\mu \partial^L_\mu A_{x\mu} .
\end{equation}
The strategy in this case \cite{Rossi:hv} is to alternate a Langevin evolution step
of order $\epsilon$ with an order $\epsilon$ gauge transformation
chosen as (\ref{GT}). That means that $\alpha$ must be rescaled with
$\epsilon$ in the real simulations.
Again one can show that such an operation introduces a force that keeps the norm 
of the gauge field limited. The iteration of (\ref{GT}) alone would drive the 
system towards a minimum of the gauge field norm
\[
N[w] = \sum_{n,\mu} \mbox{Tr} \left( A^w_{x\mu} \, A^{w \dag}_{x\mu} \right)
= \sum_{n,\mu} \mbox{Tr} \left[ (\log U^w_{x\mu}) 
(\log U^w_{x\mu})^{\dag} \right].
\]
Notice that this results in fixing a popular gauge, since such a minimum is characterized 
by the Landau condition
\[
\sum_\mu \partial^L_\mu A^a_{x\mu} \, = \, 0.
\]
There is a natural NSPT implementation of (\ref{GT}):
\[
\left( 0 < \alpha < 1 \right) \qquad 
w_{x} = \sum_{k>0} \beta^{-\frac{k}{2}} w^{(k)}_{x} \qquad 
w^{(k)}_{x} = - \alpha 
\sum_\mu \partial^L_\mu A^{(k)}_{x\mu}. 
\]
This prescription results in having also the gauge transformation implemented as a perturbative expansion. To be definite
\begin{equation}
	e^{w_{x}} = 1 + \beta^{-\frac{1}{2}} w^{(1)}_{x} + \ldots
\end{equation}
As expected, the (order by order) gauge transformation does not change the 
overall perturbative structure of the field, which is again 1 plus fluctuations 
of order $O(\beta^{-\frac{1}{2}})$. 
After the introduction of Stochastic Gauge Fixing the stochastic dynamics to 
implement is thus 
\begin{eqnarray}\label{LanGF}
	U_{x\mu}' & = & e^{-F_{x\mu}[U,\eta]} \; U_{x\mu}(n) \non \\
	U_{x\mu}(n+1) & = & e^{w_{x}[U']} \; U_{x\mu}' \; e^{-w_{x+\hat{\mu}}[U']},
\end{eqnarray}
where one should keep in mind that everything is to be implemented \emph{order by order} 
and $F$ and $w$ have to be taken as in Eq.~(\ref{eq:F}) and Eq~(\ref{GT}). Notice 
that there is a very simple way of looking at Eq~(\ref{LanGF}), which 
can be rewritten to $O(\epsilon)$ as $U_{x\mu}(n+1) = e^{-F_{x\mu}[U^G,\,G\eta G^{\dag}]} \; U_{x\mu}^G(n)$, 
$U_{x\mu}^G(n)$ being $e^{w_{x}[U]} \; U_{x\mu}(n) \; e^{-w_{x+\hat{\mu}}[U]}$. 
The effect of performing the gauge transformation after 
the Langevin step is equivalent to taking the Langevin step on a gauge equivalent configuration,  
but with a choice of $\eta$ given by 
$e^{w_{x}} \,\, \eta_{x\mu} \,\, e^{- w_x}$
(random field is gauge covariant, while equations of motion are gauge invariant). 
With this respect it is obvious why the gauge invariant quantities are not affected 
in the asymptotic limit of averaging over $\eta$. \\

Let us now proceed to understand what is the effect of Stochastic Gauge Fixing. 
This turns out to provide a force that keeps contained all the gauge
degrees of freedom of all the perturbative components of the fields. 
In fact consider again how the Langevin equation (in 
Fourier space, in continuum regularization) is modified by the 
gauge fixing term:
\ba \label{EqLangGF}
\dot{A}^a_\mu(k;t) \, &=& 
\, - k^2 (T_{\mu\nu}(k) + \frac{1}{\alpha} L_{\mu\nu}(k))\, A^a_\nu(k;t) + \\
&& + I_\mu^{(3)a}(k;t) + I_\mu^{(4)a}(k;t) + 
I_\mu^{(GF)a}(k;t) + \eta^a_\mu(k;t), \nonumber
\ea
where $I_\mu^{(GF)a}(k;t)$ is the new three gluon interaction term introduced
by gauge fixing. From the equation above one can obtain a system of 
equations as in (\ref{Lansystalg}) and obtain a formal solution which reads 
(instead of ({\ref{Sol}}))
\begin{equation} \label{SolGF}
A^{a (n)}_\mu(k;t) \,  = \,  
T_{\mu\nu} \int_0^t ds  \, e^{-  k^2 (t-s)} f^{a (n)}_\nu(k,s) + 
L_{\mu\nu} \int_0^t ds \, e^{-\frac{k^2}{\alpha} (t-s)} f^{a (n)}_\nu(k,s).
\end{equation}
One can see that now the longitudinal ($L_{\mu \nu}$) degrees of freedom
have a damping factor. We will use this observation in section 
\ref{subsproof} where we will consider the problem of the
convergence of the process. The reader is referred to the figures in \cite{DRMMOLatt94} 
to have a feeling of how effectively Stochastic Gauge Fixing 
keeps fluctuations under control.

The relation between Stochastic Gauge Fixing and the Faddeev-Popov prescription
\cite{FP} was first studied in \cite{Baulieu:1981ec}.
In section \ref{GF} we will show how it is possible to recover
the case of the Landau gauge.
\subsection{Regularization of the zero modes}
The problem of zero modes manifests itself in the formal solutions (\ref{Sol}) and 
(\ref{SolGF}). As already pointed out, the mechanism that leads to gradually diverging 
fluctuations is analogous to the one related to gauge degrees of freedom. 
Notice that in a finite lattice regularization the contribution of zero modes is finite, and particularly relevant for small lattices.
It is interesting to notice that a similar problem occurs in finite volume 
lattice perturbation theory. The generic form for a given Feynman graph is in this 
case a 
finite sum which in general entails a singularity for zero momentum. 
A common prescription consists in simply dropping the $k=0$ contribution \cite{HK85}.
This is expected to reproduce the perturbative expansion of the 
lattice regularized functional integral in the infinite volume limit. 
In order to compute the perturbative expansion exactly associated
to a lattice regularization in any finite volume one should define
a consistent prescription in both perturbative and non perturbative
context (for example with twisted boundary conditions 
\cite{LW86}\cite{GaJKa81}).
One possible choice for NSPT is to impose such a kind of boundary conditions. 
Another (simpler) approach is the subtraction of zero modes that can be obtained by
imposing the condition:
\begin{equation} \label{zmconst}
\int dx \, A^{a(n)}_\mu(x)=0.
\end{equation} 
This means that there is no zero mode contribution to the field. This should be 
regarded as a prescription on a single mode, becoming irrelevant in the infinite 
volume limit for a quantity which does not have IR problems. 
This constraint must be imposed at each evolution step
and for all perturbative orders.
In fact the Gaussian noise $\eta^a_\mu(x)$ gives a contribution to
the zero modes at each evolution step. Moreover the zero modes of higher orders take 
contributions also from non zero modes of the lower orders. For instance
\ban
A^{a(n)}_\mu(0;t) \,  &=& \,  \int_0^t ds \frac{i g f^{abc}}{2 (2 \pi)^d} +\\ 
&& +\, \sum_{m=0}^{n-1}  \int dp dq \, \delta(0+p+q) \, 
A_\nu^{b(n-1-m)}(-p;t) \, A_\sigma^{c(m)}(-q;t) \, 
v_{\mu\nu\sigma}^{(3)}(0,p,q) \, + \\
&& +\,  \mbox{4gluons} + \mbox{G F.}
\ean
To keep condition (\ref{zmconst}), we enforce it after each evolution step. 

\subsection{Convergence of the process}
\label{subsproof}

Now we come to the question of convergence of correlation functions 
of $M$ perturbative components of the fields (\ref{funzcorr}), which in momentum 
space read 
\begin{equation}\label{funzcorrk}
\langle\prod_{1\leq j \leq M} A^{(p_j)}_{\mu_j}(k_j;t) \rangle.
\end{equation}
As already pointed out, the correlation function above is more general than the perturbative 
component of an observable, which is already known to converge 
\cite{Flor}. However these are the basic elements which 
build any observable, and we want to make sure that all of them 
converge, in order to have a safe numerical algorithm. 
Notice that we are not interested in the
continuum or thermodynamic limits, at this stage. We are simply concerned 
with the existence of a limit distribution for every 
given finite lattice, where the simulations are performed.
Therefore, for the rest of this section, a finite lattice is always
understood.

The proof of convergence was given in  \cite{AdROS} in the case of
a 'zero dimensional' field theory. Since the argument is essentially
the same, here we will only stress what is peculiar for a 4-dimensional
gauge field theory.

First notice that any correlation function of free  fields
$A^{(0)}$ converges at least as $O(e^{-q^2t})$, where $q$ is the 
lowest Fourier component that contributes to the correlation. This
result may be easily obtained by using the Langevin equation
for free fields.

Let us now define the {\em total} perturbative order of 
(\ref{funzcorrk}) as $P_{tot}$ $=\sum_j p_j$.
The idea is to reduce any function
like (\ref{funzcorrk}) to a sum of correlations of free fields. 
This is done by showing that (\ref{funzcorrk}) can be written as a sum 
of correlations that  have either a strictly lower $P_{tot}$,
or the same $P_{tot}$ but  a strictly lower number of fields.

First of all we write (\ref{SolGF}) in a discretized Langevin time
$t=N\epsilon$ (and $\alpha=1$). 
\ba \label{riscrittura}
A^{(0)}_\mu(k;t) &=& e^{-k^2 \epsilon} A^{(0)}_\mu(k;t-\epsilon) +
\sqrt{\epsilon}\eta_\mu(k;t),\\
A^{(j)}_\mu(k;t) &=& e^{-k^2 \epsilon} A^{(j)}_\mu(k;t-\epsilon) +
\epsilon f^{(j)}_\mu(k;t). \nonumber
\ea 
It is worth noting that the previous expression is valid
for all degrees of freedom (and with all $k^2>0$) only because
of the gauge fixing and zero momenta subtraction described in the previous
sections.

By inserting (\ref{riscrittura}) in (\ref{funzcorrk}) and taking 
first the limit $\epsilon\rightarrow 0$ (at $t=N\epsilon$ fixed) and then
the limit $t\rightarrow\infty$ one finds:
\ba \label{miaregola}
\langle\prod_j A^{(p_j)}_{\mu_j}(k_j) \rangle &=&
\frac{1}{\sum_{m=1}^M k_m^2}
\left[
\left(
\sum_{\{1\leq i \leq M | p_i= 0\}}
\sum_{\{i < j \leq M | p_j= 0\}}
 \langle  \prod_{\{1\leq h \leq M | h \ne i , j\}} 
A^{(p_h)}_{\mu_h}(k_h) \rangle
\right)
+\right. \nonumber \\
&& \left.
\left(
\sum_{\{1\leq j \leq M | p_j\ne 0\}}
 \langle  \prod_{\{1\leq h \leq M | h \ne j\}} A^{(p_h)}_{\mu_h}(k_h)
f^{(p_h)}_{\mu_h}(k_h) \; \rangle
\right)
\right]. 
\ea
The argument to deduce (\ref{miaregola}) is given in \cite{AdROS},
for the case of a 0-dimensional field theory. 
The only difference here is that fields are now labeled by discrete momenta
(remember that we are in a finite lattice).
As a consequence the correlation functions of two fields have a decay
constant which depends on momenta and gives the momentum dependent
factor in front of the square bracket. 

By iteration of (\ref{miaregola}) $P_{tot}$ times, the initial
correlation function is reduced to a finite sum of correlation functions
of free fields. This proves that any correlation function
(\ref{funzcorrk}) has a finite limit for $t\rightarrow\infty$. 

We now come to the point of the convergence rate. 
For free fields the convergence to equilibrium may be checked
directly and it turns out to be
\[
\langle\prod_j A^{(0)}_{\mu_j}(x_j;t) \rangle =  a_\infty + a_1 e^{-q^2t}+
\sum_{q^2_j > q^2} a_j e^{-q_j^2t}
\]
where $q^2$ is the lowest Fourier component entering the correlation
function. Correlations involving higher perturbative components of
the fields also have the same exponential damping factor. However also
power corrections appear, such as
\[
t^p \, e^{-q^2 t},
\]
where $p$ is of the order of $P_{tot}$. 

For a given observable at a given perturbative order, it is therefore 
possible to estimate the
time scale in which convergence to equilibrium should occur. 
This is in practice very important. Of course one would like to know 
the size of all the moments of the distributions, 
which tells the size of the fluctuations at equilibrium. 
However the determination of such moments is not easier than determining the
perturbative coefficients themselves, and even for
the toy models studied in \cite{AdROS} only partial results could
be attained. The analysis of the fluctuations must be done for any
new observable, by means of the statistical analysis
described in \cite{AdROS}. \\

We want to stress the reassuring 
message coming from \cite{AdROS}: one does not observe in the case of LGT the 
pathological behaviour of very large (and not normally distributed) fluctuations
at high orders which we found in very simple (0-dim) models. 
In the spirit of illustrating what are the typical fluctuations one has to 
live with, in the following we perform an error analysis 
on a benchmark computation, \emph{i.e.} a large set of Wilson loops.

\subsection{Analysis of performances and simulation time}
Before analyzing the fluctuations of a particular observable, let us consider
the computer time needed for a single Langevin iteration. Together with 
autocorrelation this will provide a precise estimate of the cost of 
reaching a given precision. A quick inspection of the algorithm
suggests a dependence on the linear lattice size ($L$) and the maximum
perturbative order ($p$) as
\begin{equation}\label{eq:timing}
t_S \propto L^4 \cdot\, \frac{(p^2-p)}{2}.
\end{equation}
In fact the algorithm for pure gauge fields scales with the volume
of the lattice, and most of the time is spent in performing
order by order multiplications. This formula fits well
the timings we measured (see Table~\ref{UNQt}). 
The above considerations are of course not conclusive, since they only deal 
with crude execution times. In order to further comment on the efficiency of 
the method, we present in the following the error analysis of a large set of 
Wilson loops. In this context we will also add something about autocorrelation. 

\subsection{Wilson loops in quenched NSPT}

The results on Wilson loops we are going to present were used 
in \cite{MassTerm} to compute in Lattice Perturbation Theory the static potential, 
from which the static self--energy was in turn computed. The results for the 
Wilson loops themselves have not yet been published till now. Apart from the 
interest for reconstructing the static potential, Wilson loops have many other 
reasons to be regarded as interesting (for a discussion and a good list of references 
see for example \cite{Gunnar}). On top of that they are in a sense also 
a typical benchmark. One can find other perturbative computations of 
these quantities, either in the standard approach (even an incomplete list of 
reference should at least cite \cite{WeWeWo,HK85,Curci,Gunnar}) and by a less 
conventional technique (\cite{Trott}). To our knowledge the list of results we 
present here is the largest available at the moment.\\
The expansion (\ref{sost}) was made up to the maximal order $k=6$, so that 
we could obtain results up to $O(\beta^{-3})$. This was done on a $32^4$ lattice. 
This size was the largest which was possible to simulate at this given order 
(third, as counted in the physical expansion parameter) on the computing 
facility available to us at that time (it was an \emph{APE100} in the \emph{crate} 
configuration, with $128$ Floating Point Units (FPU's) and extended memory, for 
a total peak performance of $6.4 \times 10^{9}$ Floating Point operations per second, {\em i.e.} $6.4$ GFlops). 
Notice however that any execution time 
we report in this paper refers to the \emph{APEmille} architecture, which 
is the one currently in use. \\

Measurements are taken by following 
a common procedure in Monte~Carlo simulations. 
One first lets the system thermalize and then starts collecting 
configurations which are stored and used to measure 
various observables. Notice that in the case of NSPT this results in a 
database which asks for a big storage area. Since the fields are 
replicated in perturbative orders, one single configuration of a $32^4$ lattice 
up to sixth order amounts to $1.7$ Gbytes of binary data\footnote{The space requested 
can be substantially reduced (a factor $8/18$) if one stores the Lie algebraic fields.}. 
As already pointed out several times, configurations have to be collected at least 
at two different values of the Euler time step. An important issue is of course 
the decision on the frequency at which one should write out the configurations. 
We usually measure Wilson loops (for which 
many results are already known at least up to $g^4$ order) to decide when the system 
is effectively thermalized 
(one can check several sizes and so several effective scales). 
We also monitor on the fly at least the basic plaquette (which is anyway measured 
to compute the equations of motion entailed in $F$) in order to estimate 
the autocorrelation time at least for such a (very) local quantity. 
Some autocorrelation measurements for the plaquette on a $32^4$ lattice 
are given in Table~\ref{AutoCT}. 
We did not attempt a precise determination of the autocorrelation of each Wilson loop, since 
no long enough history is available. Instead we analysed them by the bootstrap method. 
NSPT results should in general be analysed by means of the bootstrap method as
described in \cite{AdROS}. This is necessary also to take into account
the possibility that fluctuations are not normally distributed.
As already said, however, by inspection of frequencies histograms up
to $\alpha^{10}$ it turns out that no huge deviations from normality are 
observed in the case of LGT. In fact the bootstrap analysis coincides
with the traditional one. This phenomenon is not difficult to understand
if one considers the origin of the large fluctuations in 
0-dimensional models \cite{AdROS}. \\

We collect our results for all the Wilson loops up to $16 \times16$ in 
Appendix A. Of course there are larger and larger finite sizes effects. Comparing 
results with smaller lattice sizes suggests that for loops up to the extension $6 \times 
6$ finite size effects are roughly of the order of the statistical errors. 
Deviations of the order of a few percent occur at $g^6$ for 
both loop sizes of the order of half the lattice size. 
Results were obtained out of measurements of $120$ configurations. 
By inspection of the results one can see that relative errors 
range from order $10^{-4}$ for leading order up to $10^{-3}$ for third order. 
Actually the ratios of relative errors at different orders stay much the same 
for various loop sizes. At fixed order, relative errors increase roughly linearly 
with the loop size (perimeter, actually), which is quite reasonable. 
\section{Other applications}
\subsection{Perturbative expansion for gauge-fixed observables}
\label{GF}
\begin{figure}[t]
\mbox{\epsfig{figure=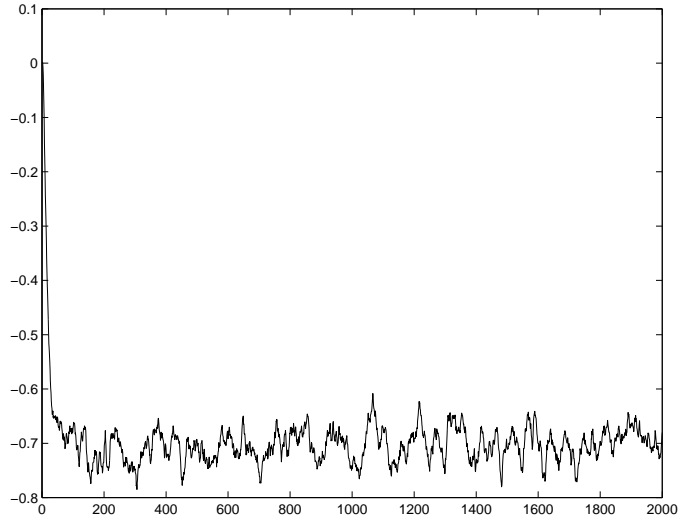,height=7.cm}}
\caption{History of the gauge dependent quantity $\mbox{Tr}U_\mu$, order $\beta^{-2}$
($4^4$ lattice).}
\label{fig:trlinka2}
\end{figure} 

Once Stochastic Gauge Fixing has entered our prescriptions for NSPT simulations, 
any correlation of gauge fields converges and also 
non gauge invariant quantities are calculable. It is interesting to inspect how 
one gets a stable signal also for non gauge invariant quantities: 
in figure~\ref{fig:trlinka2} we give an example of the evolution of a 
gauge dependent observable (the trace of the link). Even if it is not easy 
in general to make contact with the standard covariant gauges result, 
the latter can be easily recovered in the case
of Landau gauge. The strategy is the same used in standard 
(non perturbative) Monte~Carlo simulations. One first produces a thermalized 
configuration and then fixes the gauge. The Landau gauge condition is easy to attain 
if one consider Eq.~(\ref{GT}). 
\begin{figure}[b]
\mbox{\epsfig{figure=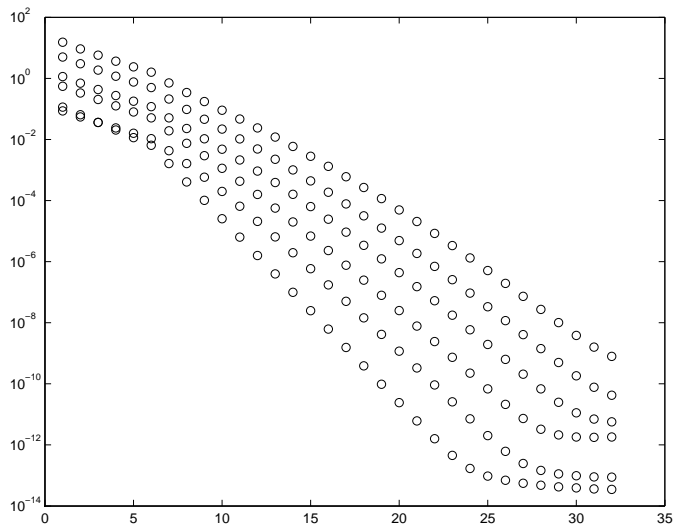,height=7.cm}}
\caption{The minimization of the quantities 
$\frac{1}{V}\sum_x \Tr \left( ({\sum_\mu \partial^L_\mu A^{(k)}_{x\mu}})^{\dag} \, 
({\sum_\mu \partial^L_\mu A^{(k)}_{x\mu}}) \right)$. Here $k=1, \ldots, 6$. 
}
\label{fig:GF}
\end{figure} 
If iterated, that gauge transformation drives the minimization 
of the norm functional, which results in turn in the Landau condition\footnote{Notice 
that this is true at every perturbative order.}. This gauge transformation is by far 
more effective if Fourier accelerated, as pointed out in \cite{Rossi:hv}. 
Figure~\ref{fig:GF} shows how the quantities 
$\frac{1}{V}\sum_x \Tr \left( ({\sum_\mu \partial^L_\mu A^{(k)}_{x\mu}})^{\dag} \, 
({\sum_\mu \partial^L_\mu A^{(k)}_{x\mu}}) \right)$ are minimized by the iteration 
of (the \emph{order by order} version of) Eq.~(\ref{GT}), 
ensuring to machine precision the Landau condition at every perturbative order. \\

We make a last point on gauge fixed computations. As it was first proposed in \cite{Lat98} 
a very appealing feature of NSPT is that there is also a way to obtain covariant 
gauges results. Without entering into details, it suffices to recall the basic formula of 
the Faddeev--Popov procedure \cite{FP}, \emph{i.e.} the complete measure in the path 
integral
\[
e^{-(S_G + S_{gf})} \; \det L[U]
\]
where one usually chooses (covariant gauges) $S_{gf}=\frac{1}{2\alpha}
\sum_{xB} (\partial^L_\mu A^B_{x\mu})^2$. 
In the previous formula a gauge fixing action has been added to the gauge (in our case, 
Wilson) action, but this, as we know, is not the end of the story: one must also keep into 
account the determinant of the Faddeev--Popov operator. Even if one can always define 
a new contribution to the action by
\[
\det L = e^{-(-\mbox{\scriptsize Tr} \ln L)}
\] 
the previous expression is in standard Perturbation Theory almost useless. One trades the 
non locality 
of this action with the introduction of the (spurious) ghost degrees of freedom. In NSPT 
one instead makes direct use of the ($\Tr \ln L$) action. Since the mechanism is just the 
same that makes it possible to treat also the fermionic determinant, we differ a few 
comments on this point to section 5.

\subsection{Perturbative expansion around non trivial vacua}
Any perturbative expansion must be performed around a fixed vacuum
configuration. One nice feature of NSPT is its flexibility 
with respect to different choices of the perturbative vacuum. 
In this very brief section we will show how, from a technical point of 
view, NSPT only needs very few modifications when the perturbative
vacuum is not the usual one (defined by $U=1$). This is not difficult to 
understand, since a vacuum configuration is basically a solution of the 
equations of motion around which the solution of Langevin equation fluctuates 
in force of the random noise. 
In order to illustrate this possibility  we will just give the relevant recipe 
in one example: the vacuum configuration
of the Schroedinger Functional scheme \cite{SF}.

In the approach of \cite{SF} the following boundary conditions are fixed for
links pointing in all space--like directions at the (boundary) 
time slices ($x^0=0$ and $x^0=T$) 
\[
U_{(\vec{x},0)k} = \exp(C), \qquad 
U_{(\vec{x},T)k} = \exp(C'),\qquad 
k=1,2,3 \qquad \forall \vec{x}.
\]
We are not interested here in the precise form of the $C$ and $C'$ (which are $SU(3)$
matrices). The classical solution with these boundary conditions is 
\begin{equation}\label{form:SFvac}
V_{x0} = 1; \qquad 
V_{xk} = \exp([x^0 C' + (T-x^0)C]).
\end{equation}

\begin{figure}[t]
\mbox{\epsfig{figure=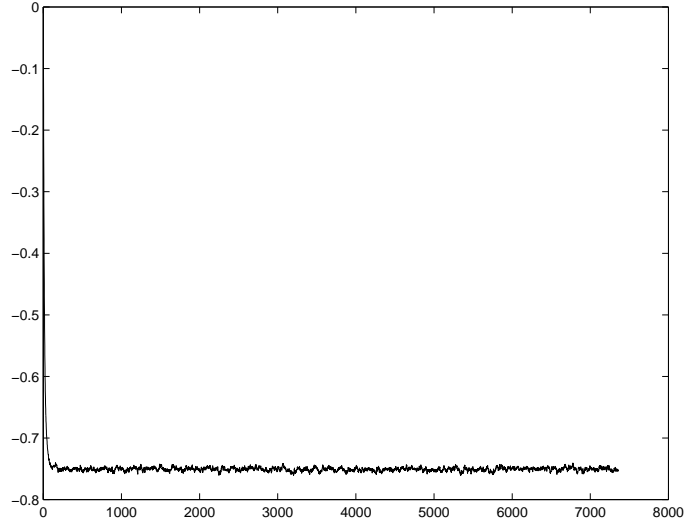,height=7.cm}}
\caption{First non trivial order ($\beta^{-1}$) of the plaquette in presence of the
vacuum configuration defined in (\ref{form:SFvac}) ($8^4$ lattice).}
\label{fig:p2}
\end{figure}

The strategy in this case simply consists in replacing (\ref{sost}) with:
\[
U_{x\mu}(t;\eta) \rightarrow V_{x\mu} + \sum_{k=1} \beta^{-k/2} U^{(k)}_{x\mu}(t;\eta).
\]
Apart from that, only slight modifications are needed. 
We are not going into details here. Figure~\ref{fig:p2} 
shows a typical signal for Schroedinger Functional NSPT (it is the leading order plaquette). 
We conclude with a last, trivial remark: it makes no difference
how complicated the vacuum configuration is. Such configuration could even be known 
only numerically, for example from a quenching procedure minimizing the equations of motion. 

\section{Inclusion of dynamical fermions}
\label{fermions}
We now come to the discussion of 
how to include the contribution coming from fermionic loops. This was first 
discussed in \cite{Lat00} and a first, preliminary discussion of the implementation 
can be found in \cite{Lat02}. 
Let us first of all define our notation. We write $M[U]$ for 
the fermion matrix (\emph{i.e.} the Dirac operator). Even if most of what follows 
is valid for a wide class of fermionic action, we will specify to the case of 
Wilson fermions (which is the case for which we will present prototype results later) 
with Wilson parameter $r=1$: 
\begin{eqnarray}
	S_F^{(W)} & = & \sum_{xy} \bar{\psi}_x \, M_{xy}[U] \, \psi_y \non \\
	& = & \sum_x (m+4) \, \bar{\psi}_x \, \psi_x - \half \sum_{x\mu} \left( 
	\bar{\psi}_{x+\hat{\mu}} \, (1+\gamma_\mu) \, U^{\dag}_{x\mu} \, \psi_x \, + \, 
	\bar{\psi}_x \, (1-\gamma_\mu) \, U_{x\mu} \, \psi_{x+\hat{\mu}}\right). \non
\end{eqnarray}
Let us also write down the explicit form of the matrix element of $M[U]$:
\begin{eqnarray} \label{eq:M_ab}
	M_{y \beta b, \, z \gamma c}[U] & = & \sum_x (m+4) \, \delta_{yz} \, \delta_{\beta\gamma} 
	\, \delta_{bc} + \\
	& & - \half \sum_{\nu=1}^{4} \left( 
	\delta_{y,z+\hat{\nu}} \, (1+\gamma_\nu)_{\beta\gamma} \, (U^{\dag}_{z\nu})_{bc} \, + \, 
	\delta_{y,z-\hat{\nu}} \, (1-\gamma_\nu)_{\beta\gamma} \, 
	(U_{z-\hat{\nu}\,\nu})_{bc} \,\right). \non
\end{eqnarray}
In the following we will drop in our notation the dependence of $M$ on $U$. 
We will also make no comment on the dependence on the number of flavors $n_f$, 
since it is trivial to work it out. Actually most of the NSPT simulations that 
we have been running involve a couple of 
degenerate massless quarks. This is the most relevant case, for example 
for the computation of renormalization constants for QCD. 
The functional integral for the fermionic degrees of freedom can be computed, 
resulting in the well known fermionic determinant contribution $\det M$. 
It is worth to recall here the point that we have already made in the previous 
section: the Faddeev---Popov action has the same determinant structure \cite{Lat98} 
and this means that the strategy for including fermions is just the same needed to implement the Faddeev--Popov procedure (without ghost fields). We will come back to 
this point later. 

Once the fermionic contribution has been integrated out, the result appears as a determinant and one has to manage a path integral which is integrated only on the gauge 
degrees of freedom, but with a new weight given by 
\begin{equation}
	e^{-S_G} \det M = e^{-S_{eff}} = e^{-(S_G-Tr \ln M)}.
\end{equation}
In the previous formula one has rewritten the determinant as a contribution to a new 
(effective) gluonic action. 
Let us see what is the effect of this on the prescription for writing 
down the Langevin equation (\ref{Langevin}). One simply needs to replace
\begin{eqnarray} \label{newLangevin}
\DLie S_G \mapsto \DLie S_{eff} & = & \DLie S_G - \DLie \Tr \ln M  \non \\
	  &  = &    \DLie S_G - \Tr ( (\DLie M) M^{-1}).
\end{eqnarray}
Eq. (\ref{newLangevin}) asks for computing the trace of a product of two matrices:
the Lie derivative of the Dirac operator and the inverse of the Dirac operator itself. 
The real difficulty does not come from the Lie derivative, which is pretty simple, 
since it is substantially local:
\begin{eqnarray}\label{DM_ab}
	\DLie M_{y \beta b, \, z \gamma c} = \;\; \frac{i}{2} & ( &
	\delta_{y,z+\hat{\mu}} \, \delta_{z,x} \, (1+\gamma_\mu)_{\beta\gamma} \, (U^{\dag}_{x\mu}T^a)_{bc} \, + \, \\
	& & - \, \delta_{y,z-\hat{\mu}} \, \delta_{x,z-\hat{\mu}} \, (1-\gamma_\mu)_{\beta\gamma} \, 
	(T^a U_{x\mu})_{bc} \,).\non
\end{eqnarray}
The difficult task comes from having to face an inverse, {\em i.e.} non--locality. Eq. (\ref{newLangevin}) and the solution given to it by the Cornell group \cite{Batrouni} were 
in a sense a prototype for fermionic simulations. The idea is to introduce another 
Gaussian source $\xi$ 
\[
\langle \xi_i \xi_j \rangle_{\xi} = \delta_{ij}
\] 
which enters the new version of Eq.~(\ref{eq:F}), which reads  
\begin{eqnarray}
F=T^a(\eps \Phi^a + \sqrt{\eps} \eta^a) \non
\end{eqnarray}
\vspace{-0.5cm}
\begin{eqnarray}\label{eq:Fnew} 
\Phi^a = \left[\DLie S_G - 
     \Real{\Dag{\xi_k}(\DLie M)_{kl}(M^{-1})_{ln}\xi_n}\right]. 
\end{eqnarray}
In the previous formula all the repeated indices are to be summed over, keeping also in mind 
that $k,l,n$ are multi--indices, to be intended as in Eq.~(\ref{eq:M_ab}). This of course means 
that the field $\xi$ has got, on top of position (or momentum), the degrees of freedom 
of a so-called \emph{spin--color} field. 
The evolution of the process will now average both on $\eta$ and on 
$\xi$. The average on $\xi$ will in particular results in 
\begin{eqnarray}
\langle \Phi^a\rangle_\xi &=& \left[\DLie S_G - 
     \Tr((\DLie M)M^{-1})\right]  \non \\
&=& \DLie \left[S_G - \Tr(\ln M) \right] \non
\end{eqnarray}
in which one can recognize Eq.~(\ref{newLangevin}). One can now rewrite 
Eq.~(\ref{eq:Fnew}) as 
\begin{equation} \label{Fpsi}
	\Phi^a = \left[\DLie S_G - 
     \Real{\Dag{\xi_l}(\DLie M)_{ln}\psi_n}\right] 
\end{equation}
where the vector $\psi$ is the solution of a linear system 
\[
M_{kl} \psi_l = \xi_k.
\]
As already pointed out, this was in a sense a prototype solution for fermionic 
simulations of Wilson fermions action. One has ended up with the inversion of a sparse matrix on a given vector. Of course the fact that the matrix $M$ is sparse 
is crucial for the actual efficiency of the method. \\
We now proceed to our 
NSPT version of the algorithm. Both Eq.~(\ref{newLangevin}) and the prescription 
entailed in Eq.~(\ref{eq:Fnew}) can be simply translated according to our \emph{order 
by order} prescription. This means first of all that the matrix $M$ itself gets expanded 
as a power series
\begin{equation}
	M = M^{(0)} + \sum_{k>0} \bek M^{(k)}.
\end{equation}
By direct inspection of Eq.~(\ref{eq:M_ab}) one realizes that the expansion of $M$ is trivial (it depends linearly on the $U_{x\mu}$): 
\begin{eqnarray} 
	M^{(k)}_{y \beta b, \, z \gamma c}[U] & = & \sum_x (m^{(k)}+4\,\delta_{k0}) 
	\, \delta_{yz} \, \delta_{\beta\gamma} 
	\, \delta_{bc} + \non \\
	& & - \half \sum_{\nu=1}^{4} \left( 
	\delta_{y,z+\hat{\nu}} \, (1+\gamma_\nu)_{\beta\gamma} \, (U^{(k) \, \dag}_{z\nu})_{bc}
	 \, + \, \delta_{y,z-\hat{\nu}} \, (1-\gamma_\nu)_{\beta\gamma} \, 
	(U^{(k)}_{z-\hat{\nu}\,\nu})_{bc} \,\right). \non
\end{eqnarray}
The only non-trivial feature of the previous expression is that also the mass acquires 
an expansion. This is due to counterterms, as we will discuss later. 
Much the same holds for $\DLie M$, as it appears from Eq.~(\ref{DM_ab}). One now has to face the inversion of $M$ as a matrix 
power expansion. 
It is easy to compute
\begin{eqnarray}
M^{-1} &=& \sum_{k=0} \bek {M^{-1}}^{(k)} \non \\
       &=& {M^{(0)}}^{-1} + \sum_{k>0} \bek {M^{-1}}^{(k)}. 
\end{eqnarray}
The notation enlightens the fact that \emph{the zeroth-order of the inverse is the inverse 
of the zeroth-order}. As for higher orders, a simple recursive relation holds: 
\begin{eqnarray}\label{eq:rec}
{M^{-1}}^{(1)} &=& - {M^{(0)}}^{-1} M^{(1)} {M^{(0)}}^{-1} \non \\
{M^{-1}}^{(2)} &=& - {M^{(0)}}^{-1} M^{(2)} {M^{(0)}}^{-1} \non \\
	       && - {M^{(0)}}^{-1} M^{(1)} {M^{-1}}^{(1)} \non \\
{M^{-1}}^{(3)} &=& - {M^{(0)}}^{-1} M^{(3)} {M^{(0)}}^{-1} \non \\
	       && - {M^{(0)}}^{-1} M^{(2)} {M^{-1}}^{(1)} \non \\
	       && - {M^{(0)}}^{-1} M^{(1)} {M^{-1}}^{(2)} \non \\
\ldots		&& \non \\
{M^{-1}}^{(n)} &=& - {M^{(0)}}^{-1} \; \sum_{j=0}^{n-1} \, M^{(n-j)} {M^{(j)}}^{-1}  \\
\ldots		&& \non 
\end{eqnarray}
We now mimic the construction in Eq.~(\ref{eq:Fnew}) introducing the new random field $\xi$. Notice that this is a field with no power expansion. Once the product $(\DLie M)\,(M^{-1})$ has got 
translated in a matrix power expansion (\emph{i.e.} it has become a matrix sum of perturbative orders), the $\xi$'s
are in charge of taking the (stochastically evaluated) trace of the various orders, 
which results in getting the power expansion of $\Tr((\DLie M)M^{-1})$. Just as in Eq.~(\ref{Fpsi}) it was useful to introduce the vector $\psi$, it is now useful to define a (power expanded) new field 
\begin{equation}
	\psi^{(j)} \equiv {M^{-1}}^{(j)} \xi.
\end{equation}
We now have a compact expression for the $n^{th}$ order of the fermionic contribution in Eq.~(\ref{Fpsi})\footnote{$k$ and $l$ are again dummy (multi-)indices over which a summation is understood.}, which reads 
\[
\sum_{j=0}^{n} \; \xi_k \left(\DLie M \right)_{kl}^{(j)} \psi_l^{(n-j)}. 
\]
Having already made the point that a simple recursive formula holds for 
Eq.~(\ref{eq:rec}), it is straightforward to notice that a similar relation 
holds for the computation of the fields $\psi^{(j)}$:
\begin{eqnarray}\label{eq:xij}
\psi^{(0)} &=& {M^{(0)}}^{-1} \xi \non \\
\psi^{(1)} &=& - {M^{(0)}}^{-1} M^{(1)} \psi^{(0)} \non \\
\psi^{(2)} &=& - {M^{(0)}}^{-1} \left[M^{(2)} \psi^{(0)} + 
M^{(1)} \psi^{(1)} \right] \non \\
\psi^{(3)} &=& - {M^{(0)}}^{-1} \left[M^{(3)} \psi^{(0)} + 
M^{(2)} \psi^{(1)} + M^{(1)} \psi^{(2)} \right] \non \\
\ldots		 \non \\
\psi^{(n)} &=& - {M^{(0)}}^{-1} \; \sum_{j=0}^{n-1} \, M^{(n-j)} \psi^{(j)} \\
\ldots		 \non 
\end{eqnarray}
\subsection{The actual implementation of unquenched NSPT}
The structure of the original (non perturbative) Eq.~(\ref{Fpsi}) is that of a scalar product: 
one starts constructing $\psi$ 
(which is the solution of a sparse system), then apply the matrix $(\DLie M)$ to it and 
finally contracts with $\xi$. The NSPT version has of course the same overall structure, provided that every operation is intended as \emph{order by order} (and 
with the caveat that the field $\xi$ has only zero order). 
We now proceed to illustrate why there is a natural and efficient way of implementing 
this program on a computer. \\

\begin{itemize}
\item
Eq.~(\ref{eq:xij}) (or Eq.~(\ref{eq:rec})) makes it clear that there is no matrix inversion 
to undertake. The only inverse matrix is the zero order ${M^{(0)}}^{-1}$, which does not depend 
on the fields and is the standard tree level Feynman propagator, diagonal in momentum 
and color space 
\[
S = \frac{(m + \half \hat{k}^2) - i\hspace{-.2em}\not\hspace{-.2em}\bar{k}}{(m + \half \hat{k}^2)^2 + \bar{k}^2}
\] 
The only problem associated to this expression is its computation at $k=0$ in case of zero bare mass. Various infrared regularizations
are available: leave a small mass, introduce anti-periodic boundary 
conditions for
the fermionic fields in time direction or constrain the zero mode 
degree of freedom to zero.
To produce the results in the present work the last choice was adopted. 
We checked stability of results with respect to the use of a small mass. 
More on this in section 6.2 \\

\item
Notice that Eq.~(\ref{eq:xij}) suggests the construction of the various orders $\psi^{(n)}$ 
in a sequential way. At every order only one application of ${M^{(0)}}^{-1}$ 
is needed and the propagator operates on a sum of already computed quantities (\emph{i.e.} 
lower order $\psi$'s). While 
${M^{(0)}}^{-1}$ is diagonal in momentum space, all the operators needed to construct this 
sum (\emph{i.e.} the various orders of $M$) are almost diagonal in configuration space. This 
suggest the strategy of going back and forth from Fourier space. \\

\item
This structure is common to many fermionic regularizations. This is also the 
right point to go back to the Faddeev--Popov determinant. In the latter case the 
zero order inverse is again reminiscent of a tree level propagator, but this time a scalar 
one\footnote{This does not come as a surprise: the determinant involves the degrees of freedom 
which in standard perturbative computations are absorbed by ghosts.}. As for higher orders, 
these are in the case of Faddeev--Popov determinant naturally expressed in the adjoint 
representation. The strategy of going back and forth from Fourier space stays much the same. \\

\item
Previous considerations clearly stress the need for a reasonably efficient Fast Fourier Transform (\emph{FFT}). 
Apart from the stochastic dynamics, actually momentum space is the natural stage also 
for the measurement of observables which are diagonal in that space (think about quark 
bilinears). \\

\item
By inspecting Eq.~(\ref{eq:Fnew}) one can not recognize whether the effective time scales 
associated to the gauge and (pseudo)fermions drifts are the same. This is a very general 
and well known point \cite{SextWein}. Besides this observation, even if one decides to 
make use of the same time step for both drifts, it is anyway useful to take advantage of 
the fact that one can always live with $O(\eps)$ errors. In our Euler scheme this is an effect which has to be in any case extrapolated to zero. As a matter of fact 
the computations of the gauge and (pseudo)fermions contribution to the drift are quite 
different from the point of view of the actual implementation on a computer. This suggests 
to break down the evolution step in the following way: \\

\begin{itemize}
\item
Evolution by the pure gauge contribution to the drift $F_{gauge}=\eps \, \nabla S_G + 
\sqrt{\eps} \, \eta$. \\

\item
Construction of the $\psi^{(n)}$; this is the only non--local piece of the 
computation, the non--locality being anyway traded for multiple 
applications of an {\em FFT}, after which every operation is indeed local. \\

\item
Evolution by the fermionic contribution to the drift $F_{ferm} = \eps \, \Real{\Dag{\xi}(\nabla  M)\psi}$. 
This does not present any structural difference with respect to the first module. \\
\end{itemize} 
One should always keep in mind that also a step of Stochastic Gauge Fixing has to be 
taken at the end of this sequence. \\
\end{itemize} 

\noindent There is a last point to be made concerning the NSPT expansion of the unquenched 
Langevin equation. In inspecting the structure of 
\[
F =  \eps (\nabla S_G - \nabla \, \Tr \ln M) + \sqrt{\eps} \, \eta  
\]
one should keep in mind our rescaling of the time step 
$\epsilon'=\epsilon\beta$ which was meant to get a consistent 
perturbative expansion. Notice that this leaves the fermionic contribution to 
the drift with an overall $\beta^{-1}$ in front. This does not come as a surprise. 
Once the fermionic degrees of freedom have been integrated, one is left only 
with the gauge bosons. In the equation of motion for the gluons fermionic 
contributions enter as loops and to "dress" a gluon one needs at least an order 
$\beta^{-1}$. 

\subsection{Analysis of simulation times}
We saw in Sec.~3.4 that the simulation time for the pure gauge implementation of 
NSPT is in good agreement with the expectations: it scales with 
the volume and is dominated by the \emph{order by order} multiplications (see 
Eq.~(\ref{eq:timing})). Since the unquenched version of the method introduces 
substantial changes, it is of course compelling to verify what is the impact on 
the simulation times. \\
\begin{table}[t]
\caption{Simulation times in (seconds/iteration $\times$ number of 
processors). Run in $8^4$ lattice
were performed on an \emph{APEmille board}; $16^4$ on an \emph{APEmille unit}; 
$32^4$ on an \emph{APEmille crate}. 
Absolute values in seconds
are only for illustration, being strongly implementation dependent. One 
should observe in particular
the scaling in $L$, in $p$ and compare Quenched with Unquenched.\label{UNQt}}
\begin{center}
\begin{tabular}{|c|c|c|c|}
\hline
\, & \, & \, & \, \\ 
lattice size $L$  & order $p$ & $n_f = 0$ & $n_f \neq 0$ \\
\hline
\hline
$8$ & $g^6$                       & 27.8 & 48.0 \\
$8$ & $g^8$                       & 49.0 & 81.6 \\
$8$ & $g^{10}$                    & 78.7 & 128.7 \\
\hline
$16$ & $g^6$                      & 453  & 814 \\
\hline
$32$ & $g^6$                      & 7168  & 12979 \\
\hline
\end{tabular}\\
\end{center}
\end{table}
This is a good point to comment on our own implementation of NSPT programs. The main 
NSPT project for Lattice QCD is run on the \emph{APE} architecture. The first 
implementation was that of the quenched version on the \emph{APE100} family. 
The unquenched version has been 
developed on \emph{APEmille}. Our $FFT$ implementation mimic \cite{FFT}, which is 
based on a $1$--dim $FFT$ plus transpositions. The latter operation is the one asking 
for local addressing on a parallel architecture, which made it necessary to wait 
for \emph{APEmille} in order to implement unquenched NSPT on \emph{APE}. \cite{Lat02} collects some other comments on our \emph{APEmille} implementation. 
In the last two years the growth in the computational power 
available on \emph{PC}'s made it worth to develop a $C^{++}$ implementation which 
is now run on medium size \emph{PC}--clusters, usually to assess finite volume effects 
(we simulate small lattices on \emph{PC}'s and large lattices 
on \emph{APE}). In Table~\ref{UNQt} we make use of timings taken on 
\emph{APE} in order to 
assess what is the overhead in moving from quenched to unquenched NSPT\footnote{The 
quarks involved were degenerate in mass, so that the dependence on $n_f$ is trivial.}. 
We report the execution times for a single iteration times the number of 
processors (formally, a theoretical execution time on a single processor). 
We are mainly concerned in scaling properties. 
On a given volume (which in the example of the table is a modest $8^4$), one can 
inspect the growth of execution times as the perturbative order grows. Both in 
the column of the quenched and in the column of the unquenched simulations, the 
scaling of computational time is again consistent with the fact that order by order 
multiplications are the dominant operations. 
On each row this results in a growth in time due to unquenching 
which is roughly consistent with a factor $5/3$. One then wants to understand 
the dependence on the volume, which is the critical one, given the presence of the 
determinant: this is exactly the growth which has to be tamed by the $FFT$. 
One then compares execution times at a given order on $L=8$, $L=16$ and $L=32$ 
lattice sizes. 
Notice the different \emph{APEmille} configurations: $L=8$ is simulated on an 
\emph{APEmille board} (for a total of $8$ FPU's), while $L=16$ on an \emph{APEmille 
unit} ($32$ FPU's) and $L=32$ on an \emph{APEmille 
crate} ($128$ FPU's). One easily understands that 
$FFT$ is doing its job: the simulation time scales with the volume also for 
the unquenched version of NSPT. The very same emerges also from the ($C^{++}$) 
\emph{PC}--cluster implementation of the programs.

\begin{table}[t]
\caption{Autocorrelation times for the basic plaquette expressed in iteration number. 
\label{AutoCT}}
\begin{center}
\begin{tabular}{|c|c|c|c|c|c|}
\hline
\, & \, & \, & \, & \, & \, \\ 
lattice size $L$  &  $n_f$ & Euler time step &  $\beta^{-1}$ &  $\beta^{-2}$ & $\beta^{-3}$ \\
\hline
\hline
$32$ & $0$ & 0.01 & $\sim 50$ & $\sim 70$ & $\sim 100$ \\
\hline
$32$ & $2$ & 0.005 & $\sim 100$ & $\sim 120$ & $\sim 150$ \\
\hline
\end{tabular}\\
\end{center}
\end{table}
As already pointed out for 
the quenched case, at this level one has only compared 
crude execution times. In our experience the optimization 
of the signal to noise ratio asks for smaller values of the Euler time 
step. We halve the values of time step with respect to the quenched case, which 
for the computational overhead results 
in an overall factor $2$ on top of the $5/3$ coming from the crude execution times 
measurements. This is anyway a good message: we are talking of a difference with 
respect to the quenched case which is a factor and not an order of magnitude. 
Table~\ref{AutoCT} contains estimates of autocorrelation times for the basic plaquette 
both in quenched and unquenched case. If one rescales the values keeping 
into account the time step, the unquenched autocorrelation time appears slightly shorter. 
Even if this is not conclusive, it is not unreasonable, since the new random field $\xi$ 
can give an extra contribution to decorrelate two successive configurations. 

A more complete assessment will be possible after having 
inspected the errors in a benchmark computation which is the same also reported  
for the quenched case, \emph{i.e.} the computation of Wilson loops.

\section{Results in unquenched NSPT}
We collect, in the following, a couple of prototype applications of unquenched NSPT: 
the computations of unquenched Wilson loops and of the (Wilson fermions) critical 
mass (to be introduced later). Both are presented to order $\beta^{-3}$ for the case 
of Wilson gauge action coupled to two mass degenerate Wilson fermions. The quark 
masses are put to zero by plugging in the relevant counterterms (more 
on this in the following). Actual computations were performed on a $32^4$ lattice 
on an \emph{APEmille} machine in the \emph{crate} configuration (128 processors for 
a peak performance of 64 GFlops). 

\subsection{Unquenched ($n_f=2$) Wilson loops}

In Appendix B we report a table of Wilson loops which are just the same as in 
Appendix A, but in the unquenched case of two massless Wilson 
quarks. As in the quenched case, the major physical motivation was the 
computation of the static potential and the static self--energy \cite{MassTermUNQ}. 
The configurations on which measurements were taken were $200$\footnote{These 
configurations are part of a wider database containing also other values of $n_f$.} 
(again, at two different values for the Euler time step). By inspection, the 
general picture for relative errors does not change with respect to the quenched 
case.  

\subsection{The $n_f=2$ critical mass for Wilson fermions to the third loop}

It is well known that the Wilson lattice regularization of fermions breaks 
chiral symmetry. This comes from the irrelevant term which enters the 
Dirac operator in order to cure the so called doubling problem. The first net 
effect is then the appearance of an additive mass renormalization, which is 
usually referred to as the critical mass. \\

Let us set up our notations. We write the two points vertex function (the inverse of the quark propagator) as 
\begin{eqnarray}
\Gamma_2(p^2,m) &=& S(p^2,m)^{-1} \nonumber \\
&=& i\hspace{-.2em}\not\hspace{-.2em}p + m - \Sigma(p^2,m) 
\end{eqnarray}
where
\begin{eqnarray}
\label{SelfE}
\Sigma(p^2,m) = \Sigma_c + m\:\Sigma_S(p^2,m) + i\hspace{-.2em}\not\hspace{-.2em}p\:\Sigma_V(p^2,m). 
\end{eqnarray}
The previous equations are written in the continuum limit. 
$\Sigma_c$ is our notation for the critical mass. In our simulations $m=0$, so that 
$\Sigma_c$ is the only contribution along the (Dirac) identity operator. For $a\neq0$, 
one has\footnote{Also the tree level mass for $a\neq0$ has an 
irrelevant contribution due to the Wilson prescription to eliminate doublers:
$\;\;\;m \mapsto m + m_W(ap)$. To shorten the notation in the following we will 
write $\Sigma_c(pa)=\Sigma_c+\tilde{\Sigma}_c(pa)$, \emph{i.e.} $\Sigma_c$ 
will be both the function of $pa$ and its value at $pa=0$, which is the 
relevant, although divergent, physical quantity.} 
$\Sigma_c \mapsto \Sigma_c+\tilde{\Sigma}_c(pa)$. 
By restoring physical dimensions one can inspect the $a^{-1}$ divergence in $\Sigma_c$. 
Because of the power divergence, the perturbative 
evaluation of the critical mass is not supposed to be good. Therefore
this quantity was in a sense a prototype for non perturbative determination of 
renormalization constants (an additive renormalization, in this case). Still, its 
perturbative computation has become sort of a benchmark. Two loops results are 
available \cite{Haris,Pelo} so that we have the chance both to check the first 
and second order and to try to understand what is the effect of the third loop 
on the (poor) convergence properties of the series. \\

\begin{figure}[t]
\mbox{\epsfig{figure=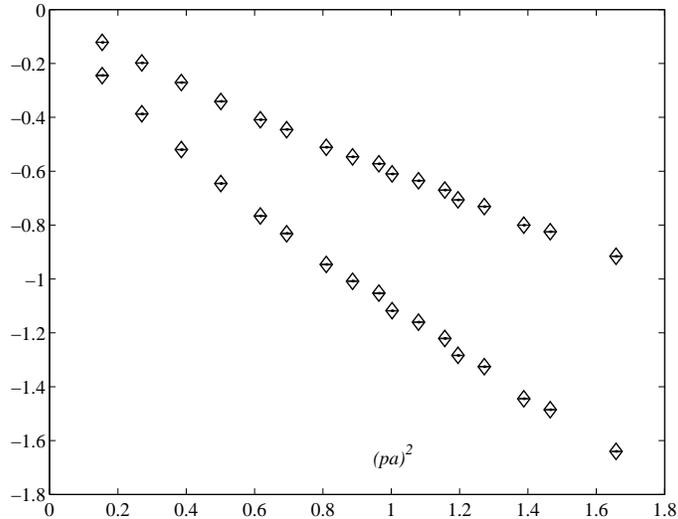,height=7.cm}}
\caption{First and second loop of $\Sigma_c$ after the relevant counterterms 
have been plugged in. Diamonds are the fitted points. Error bars on the 
measures are hard to distinguish.}
\label{CM12}
\end{figure}

\begin{figure}[b]
\mbox{\epsfig{figure=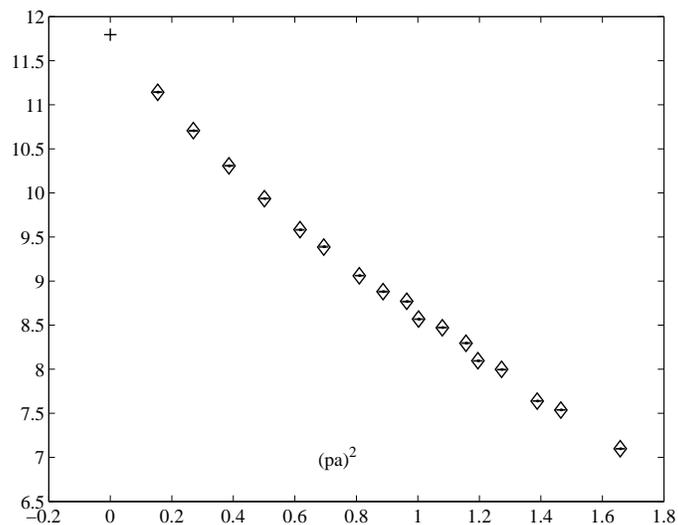,height=7.cm}}
\caption{The third order critical mass $\Sigma_c^{(3)}$.
As in Figure~\ref{CM12}, diamonds are the fitted points (error bars are again very tiny). 
The point marked as a cross is the extrapolated value.}
\label{CM3}
\end{figure}

The computation is performed on sets ranging from $200$ to $60$ configurations, 
depending on the value of momentum\footnote{This means that we have not evaluated 
the propagator on all the $200$ configurations for all the momenta that are plotted in figures \ref{CM12} and \ref{CM3}. As it can be inspected, this does not appear 
manifest in the end, because the quality of errors is quite good.}. These are part 
of the same configurations database on which also Wilson loops were measured.
As already said, the mass of the two quarks was put to zero by plugging in the 
relevant counterterms. This is a good point to comment on what this means. Once 
one realizes that there is an additive renormalization entering the stage at one 
loop, each new result for this additive contribution to the mass has to be taken 
into account in the computation of higher loops. In conventional, 
diagrammatic Perturbation Theory this means to take into account new effective 
vertices (counterterms, actually). NSPT is not different with this respect. The 
higher the loop order we want to go, the more counterterms we have to plug in 
coming from lower loops. 
Since we know the first and second counterterms from \cite{Haris,Pelo}, there is no 
problem in trying to compute the third loop. Also the latter should in turn be plugged in 
if one wanted to compute the fourth order correction. Notice that plugging the 
counterterms in results in what one would call a renormalized perturbation theory 
(with respect to the additive mass renormalization). Since counterterms are taken into 
account, the result that we obtain for the first and second order of the critical mass 
is zero, as can be seen from figure~\ref{CM12}. This is the way to check that we 
agree with previous computations at first and second loop level. \\

In order to compute $\Sigma_c$ we first compute the propagator and then invert it to 
obtain $\Gamma_2$. From $\Gamma_2$ the critical mass is obtained by taking the 
projection along the (Dirac) identity operator. 
The computation is performed in the Landau gauge by using the 
gauge fixing procedure that was explained in section~\ref{GF}. The computation of a 
propagator is the prototype measurement in the fermionic sector of QCD. We know 
from the Wick theorem that every observable reduces to a combination of inverses of 
the Dirac operator. 
This is the case in which the measurement we want is the propagator itself. 
In NSPT we measure it in 
momentum space. Since the configuration average will be diagonal in momentum, it 
suffices to compute on each configuration only diagonal entries of the inverse of $M$. 
Of course we still make use of our recursive relations given in Eq.~(\ref{eq:rec}) 
to compute the inverse of $M$. Actually, we make use of the recursion given in 
Eq.~(\ref{eq:xij}), since the various entries of the inverse matrix are obtained by 
operating on sources vectors $\xi$ which are now (Kroneker) $\delta$--functions. 
This is the trivial observation that the $A_{ij}$ entry of the matrix $A$ can be 
computed as a scalar product 
\[
A_{ij} = \sum_{k,l} \xi^{(i)}_k A_{kl} \xi^{(j)}_l
\]
where the entries of the vector $\xi^{(a)}$ are defined as $\xi^{(a)}_k = \delta_{ak}$. 
We make use of the symmetry properties of the propagator by averaging over 
the momenta that are connected by hypercubic group transformations. This is not the 
only reason why the hypercubic symmetry is important. As a matter of fact, in the 
continuum limit one recovers the $O(4)$ Euclidean symmetry, \emph{i.e.} $\Sigma_V$ 
and  $\Sigma_S$ are functions of $p^2$ (logarithms, actually). 
At finite values of the lattice spacing only hypercubic symmetry 
is there and as a consequence the $p$--expansion for a generic entry of the propagator 
contains components on all the hypercubic invariants one can construct from powers of 
$p$. By 
restoring physical dimensions one realizes that the expansions are in power of $ap$. 
This means that we have a handle on lattice spacing effects and this is the way to look at 
figures \ref{CM12} and \ref{CM3}. We write the expansion of the critical mass as 
\begin{equation}
- \Sigma_c = \Sigma_c^{(1)}\beta^{-1} + \Sigma_c^{(2)}\beta^{-2} + \Sigma_c^{(3)}\beta^{-3} + \ldots.
\end{equation}
Once again, what we plot in the figures are not the coefficients of this 
expansion, but those of 
$- \Sigma_c - \Sigma_c^{(1)}\beta^{-1} - \Sigma_c^{(2)}\beta^{-2}$. These coefficients  
are zero at first and second order (see figure~\ref{CM12}), while $\Sigma_c^{(3)}$ is 
the first non zero coefficient, \emph{i.e.} the one we are interested in 
(see figure~\ref{CM3}). Notice that these coefficients are plotted versus $(ap)^2$, but, 
due to the presence of higher orders invariants, they do not appear smooth. Since 
we are interested in the continuum limit of the computation, what we have to do is 
to extrapolate the curves to $0$. This is done by fitting the data taking into account 
the hypercubic invariants made of powers of $ap$. The curves in figure~\ref{CM12} 
actually extrapolates to numbers of order $10^{-3}$ and $10^{-2}$. We explicitly 
checked at one loop level that within errors we exactly reproduce the 
finite size effects expected on a $32^4$ lattice. 
The goodness of the extrapolation to zero of the first two 
coefficients
(with counterterms taken from an infinite volume computation \cite{Haris,Pelo})
is indeed a good indication on finite volume effects and sensitivity to 
our choice of infrared regularization (see section 5). \\

The final result for the third loop of the critical mass for $n_f=2$ is 
\begin{equation}
	\Sigma_c^{(3)} = 11.79^{(+2)}_{(-5)}. 
\end{equation}
The errors are estimated by looking at the stability of the fitted result with respect to varying the number of points included in the fit and the higher orders lattice invariants taken into account. They are only statistical errors, but the finite size effects are 
expected to be small.\\

In order to inspect the convergence properties of the series (which, as we have 
already said, are supposed to be poor) one could now try to repeat the analysis 
contained in \cite{Haris}, \emph{i.e.} the comparison of a fixed order in 
Perturbation Theory with the non perturbative determination of the critical mass. 
If one for examples tries to perform this little exercise at $\beta=5.6$ using for the 
non perturbative determination of the critical mass the data in \cite{Haris}, the 
ratio of the perturbative result to the non perturbative result raises from 
$0.55$ at first loop to $0.71$ at second loop and finally to $0.79$ at third loop level.  
As one knows, results can improve quite a lot moving to boosted perturbation theory. 
We do not perform this exercise for the critical mass and leave it for the reader, 
wanting to stress that this is not physically relevant in this case. Still, we want to 
stress that what we have just presented is only a prototype computation. Actually, 
by considering the projection of $\Gamma_2$ along the $\gamma$--matrices, one has 
access to the field renormalization constant. Preliminary results have been reported in 
\cite{LAT04QCD} and the complete computation of the latter will
be reported together with that of renormalization constants for quark bilinears. 
For those logarithmically 
divergent quantities a three loop computation would be certainly useful. 
\section{Conclusions}
We reviewed the application of Numerical Stochastic Perturbation Theory to 
Lattice Gauge Theories and discussed the implementation
of the method in a Monte Carlo algorithm. We studied the underlying
stochastic process and its properties of convergence. In order to assess 
the efficiency of the method we measured crude execution times and studied 
the size of the fluctuations for benchmark computations. 
After touching the subject of gauge fixed computations and expansions around 
non trivial vacua, we discussed the inclusion of dynamical fermions. We stressed 
that in this perturbative approach they are not so computationally expensive as 
in the non-perturbative case. 
We presented as benchmark unquenched computations a large set of Wilson loops and the 
critical mass of Wilson fermions to the third loop.
\section*{Acknowledgments}
NSPT comes out of a Parma--Milan collaboration. The idea of a numerical 
implementation of Stochastic Perturbation Theory on a computer was first pursued 
by G. Marchesini and 
E. Onofri, from whom the authors learnt a lot. The authors are also indebted to 
their colleagues and friends G. Burgio and M. Pepe for many valuable discussions. 
We also would like to acknowledge A.S.~Kronfeld for suggestions. 
In the last two years younger collaborators joined the little group working on 
NSPT: A.~Mantovi, V.~Miccio and C.~Torrero, to whom we are also indebted for their 
precious contribution. 
Much of the numerical work has been performed
on the APE machines the Milan--Parma group was endowed of by I.N.F.N.. 
F.D.R. acknowledges support from both Italian MURST 
under contract 2001021158 and from I.N.F.N. under {\sl i.s. MI11}.
L.S. acknowledges support from DFG through the 
Sonderforschungsbereich `Computational Particle Physics' (SFB/TR 9).
\newpage
\begin{appendix}
\section{Quenched Wilson Loops}

\begin{table*}[h]
	\begin{center}
\begin{tabular}{|c|c|c|c|}
\hline
\, & \, & \, & \\
Loop dimensions  & order $\beta^{-1}$ & order $\beta^{-2}$ & order $\beta^{-3}$ \\
\hline
\hline
$1 \times 1$ &  -2.0000(2) & -1.2205(6) & -2.9572(34) \\
\hline
$1 \times 2$ &  -3.4488(4) & -0.1380(13) &  -2.2088(59) \\
\hline
$1 \times 3$ &  -4.8085(6) &   2.7795(23) &  -0.5943(72) \\
\hline
$1 \times 4$ & -6.1503(8) & 7.4823(38) & -0.5589(96) \\
\hline
$1 \times 5$ &  -7.4872(10) & 13.961(6) & -4.460(16) \\
\hline
$1 \times 6$ &  -8.8225(13) & 22.217(9) & -14.684(27) \\
\hline
$1 \times 7$ &  -10.157(2) & 32.247(14) & -33.572(51) \\
\hline
$1 \times 8$ &  -11.491(2) & 44.054(19) & -63.501(81) \\
\hline
$1 \times 9$ &  -12.825(2) & 57.637(26) & -106.84(12) \\
\hline
$1 \times 10$ &  -14.159(3) & 73.001(34) & -166.05(19) \\
\hline
$1 \times 11$ &  -15.492(3) & 90.135(44) & -243.38(27) \\
\hline
$1 \times 12$ &  -16.825(4) & 109.05(6) & -341.19(39) \\
\hline
$1 \times 13$ &  -18.158(4) & 129.73(7) & -461.74(53) \\
\hline
$1 \times 14$ &  -19.491(5) & 152.19(8) & -607.52(71) \\
\hline
$1 \times 15$ &  -20.825(5) & 176.42(10) & -780.90(97) \\
\hline
$1 \times 16$ &  -22.158(6) & 202.44(12) & -984.2(1.3) \\
\hline
$2 \times 2$ &  -5.4770(8) & 4.3363(32) & 0.0394(94) \\
\hline
$2 \times 3$ &  -7.2455(11) &  11.630(6) & -1.226(14) \\
\hline
$2 \times 4$ &  -8.9557(15) & 21.711(10) & -10.880(27) \\
\hline
$2 \times 5$ &  -10.649(2) & 34.596(15) & -33.645(55) \\
\hline
$2 \times 6$ &  -12.335(2) & 50.290(22) & -74.23(10) \\
\hline
$2 \times 7$ &  -14.019(3) & 68.790(31) & -137.28(16) \\
\hline
$2 \times 8$ &  -15.701(3) & 90.107(43) & -227.64(26) \\
\hline
$2 \times 9$ &  -17.382(4) & 114.24(6) & -349.97(38) \\
\hline
$2 \times 10$ &  -19.062(4) & 141.19(7) & -509.09(55) \\
\hline
$2 \times 11$ &  -20.742(5) & 170.95(9) & -709.63(76) \\
\hline
$2 \times 12$ &  -22.421(6) & 203.51(12) & -956.07(1.04) \\
\hline
$2 \times 13$ &  -24.100(6) & 238.89(14) & -1253.4(1.4) \\
\hline
$2 \times 14$ &  -25.779(7) & 277.08(17) & -1606.2(1.9) \\
\hline
$2 \times 15$ &  -27.457(8) & 318.09(21) & -2019.4(2.4) \\
\hline
$2 \times 16$ &  -29.136(9) & 361.93(25) & -2497.8(3.1) \\
\hline
$3 \times 3$ &  -9.2216(18) & 23.225(12) & -12.261(33) \\
\hline
$3 \times 4$ &  -11.088(2) & 37.851(19) & -39.256(68) \\
\hline
$3 \times 5$ &  -12.919(3) & 55.637(29) & -87.89(12) \\
\hline
$3 \times 6$ &  -14.737(3) & 76.632(40) & -164.05(19) \\
\hline
$3 \times 7$ &  -16.549(4) & 100.85(6) & -273.37(30) \\
\hline
$3 \times 8$ &  -18.358(4) & 128.30(7) & -421.87(46) \\
\hline
$3 \times 9$ &  -20.164(5) & 158.99(9) & -615.46(67) \\
\hline
$3 \times 10$ &  -21.969(6) & 192.92(12) & -859.71(95) \\
\hline
$3 \times 11$ &  -23.773(7) & 230.09(14) & -1160.6(1.3) \\
\hline
$3 \times 12$ &  -25.576(8) & 270.49(18) & -1524.0(1.8) \\
\hline
$3 \times 13$ &  -27.379(8) & 314.12(21) & -1955.6(2.3) \\
\hline
\end{tabular}\\
	\end{center}
\end{table*}

\begin{table*}[p]
	\begin{center}
\begin{tabular}{|c|c|c|c|}
\hline
\, & \, & \, & \\
Loop dimensions  & order $\beta^{-1}$ & order $\beta^{-2}$ & order $\beta^{-3}$ \\
\hline
\hline
$3 \times 14$ &  -29.181(9) & 361.00(25) & -2461.2(2.9) \\
\hline
$3 \times 15$ &  -30.984(10) & 411.12(30) & -3046.9(3.7) \\
\hline
$3 \times 16$ &  -32.785(12) & 464.49(35) & -3718.3(4.7) \\
\hline
$4 \times 4$ &  -13.051(3) & 56.867(31) & -91.10(13) \\
\hline
$4 \times 5$ &  -14.960(4) & 79.111(45) & -172.59(22) \\
\hline
$4 \times 6$ &  -16.848(4) & 104.72(6) & -290.32(35) \\
\hline
$4 \times 7$ &  -18.725(5) & 133.74(8) & -450.58(54) \\
\hline
$4 \times 8$ &  -20.597(6) & 166.20(10) & -659.90(76) \\
\hline
$4 \times 9$ &  -22.465(7) & 202.09(13) & -924.5(1.1) \\
\hline
$4 \times 10$ &  -24.330(9) & 241.41(16) & -1250.7(1.5) \\
\hline
$4 \times 11$ &  -26.194(10) & 284.18(19) & -1645.3(2.0) \\
\hline
$4 \times 12$ &  -28.058(11) & 330.41(23) & -2114.5(2.6) \\
\hline
$4 \times 13$ &  -29.918(11) & 380.09(28) & -2665.0(3.3) \\
\hline
$4 \times 14$ &  -31.779(12) & 433.19(32) & -3302.6(4.2) \\
\hline
$4 \times 15$ &  -33.640(13) & 489.77(38) & -4034.5(5.2) \\
\hline
$4 \times 16$ &  -35.501(13) & 549.80(44) & -4866.5(6.5) \\
\hline
$5 \times 5$ &  -16.926(5) & 105.72(7) & -294.53(39) \\
\hline
$5 \times 6$ &  -18.860(5) & 135.72(9) & -460.82(57) \\
\hline
$5 \times 7$ &  -20.778(6) & 169.20(11) & -678.29(83) \\
\hline
$5 \times 8$ &  -22.690(7) & 206.24(13) & -954.0(1.1) \\
\hline
$5 \times 9$ &  -24.596(8) & 246.81(16) & -1294.0(1.6) \\
\hline
$5 \times 10$ &  -26.498(9) & 290.93(20) & -1705.6(2.1) \\
\hline
$5 \times 11$ &  -28.398(9) & 338.61(24) & -2195.7(2.7) \\
\hline
$5 \times 12$ &  -30.296(11) & 389.88(28) & -2771.1(3.5) \\
\hline
$5 \times 13$ &  -32.192(12) & 444.71(33) & -3438.0(4.4) \\
\hline
$5 \times 14$ &  -34.088(13) & 503.06(38) & -4202.8(5.4) \\
\hline
$5 \times 15$ &  -35.982(14) & 565.02(44) & -5073.8(6.7) \\
\hline
$5 \times 16$ &  -37.878(14) & 630.53(51) & -6056.5(8.3) \\
\hline
$6 \times 6$ &  -20.830(6) & 170.06(11) & -683.57(86) \\
\hline
$6 \times 7$ &  -22.781(7) & 207.92(14) & -965.7(1.2) \\
\hline
$6 \times 8$ &  -24.722(8) & 249.38(17) & -1314.9(1.5) \\
\hline
$6 \times 9$ &  -26.654(9) & 294.41(20) & -1737.2(2.1) \\
\hline
$6 \times 10$ &  -28.582(10) & 343.08(25) & -2240.2(2.9) \\
\hline
$6 \times 11$ &  -30.506(10) & 395.38(29) & -2831.0(3.6) \\
\hline
$6 \times 12$ &  -32.429(12) & 451.33(34) & -3516.7(4.4) \\
\hline
$6 \times 13$ &  -34.350(13) & 510.92(39) & -4303.8(5.4) \\
\hline
$6 \times 14$ &  -36.269(14) & 574.13(46) & -5198.8(6.8) \\
\hline
$6 \times 15$ &  -38.188(16) & 641.05(52) & -6211.6(8.3) \\
\hline
$6 \times 16$ &  -40.104(16) & 711.56(60) & -7345(10) \\
\hline
$7 \times 7$ &  -24.759(8) & 250.11(18) & -1320.6(1.7) \\
\hline
$7 \times 8$ &  -26.722(9) & 295.93(21) & -1751.1(2.1) \\
\hline
$7 \times 9$ &  -28.674(10) & 345.32(25) & -2262.4(2.9) \\
\hline
$7 \times 10$ &  -30.622(11) & 398.42(31) & -2863.8(3.8) \\
\hline
\end{tabular}\\
	\end{center}
\end{table*}

\begin{table*}[p]
	\begin{center}
\begin{tabular}{|c|c|c|c|}
\hline
\, & \, & \, & \\
Loop dimensions  & order $\beta^{-1}$ & order $\beta^{-2}$ & order $\beta^{-3}$ \\
\hline
\hline
$7 \times 11$ &  -32.566(12) & 455.16(36) & -3561.9(4.7) \\
\hline
$7 \times 12$ &  -34.504(13) & 515.62(41) &  -4364.0(5.8) \\
\hline
$7 \times 13$ &  -36.442(14) & 579.76(47) & -5276.8(7.0) \\
\hline
$7 \times 14$ &  -38.379(15) & 647.56(54) & -6306.9(8.7) \\
\hline
$7 \times 15$ &  -40.316(16) & 719.11(61) & -7464(10) \\
\hline
$7 \times 16$ &  -42.247(18) & 794.28(71) & -8751(12) \\
\hline
$8 \times 8$ &  -28.706(9) & 346.09(25) & -2270.8(2.7) \\
\hline
$8 \times 9$ &  -30.676(10) &  399.80(29) & -2879.5(3.6) \\
\hline
$8 \times 10$ &  -32.640(12) & 457.27(35) & -3587.5(4.7) \\
\hline
$8 \times 11$ &  -34.596(12) & 518.38(40) & -4400.3(5.7) \\
\hline
$8 \times 12$ &  -36.551(14) & 583.23(46) & -5325.6(6.9) \\
\hline
$8 \times 13$ &  -38.503(15) & 651.77(53) & -6370.2(8.4) \\
\hline
$8 \times 14$ &  -40.451(16) & 724.04(60) & -7542(10) \\
\hline
$8 \times 15$ &  -42.401(18) & 800.10(68) & -8850(12) \\
\hline
$8 \times 16$ &  -44.346(19) & 879.77(77) & -10296(14) \\
\hline
$9 \times 9$ &  -32.660(12) & 457.82(35) & -3593.5(4.8) \\
\hline
$9 \times 10$ &  -34.639(13) & 519.60(42) & -4415.1(6.1) \\
\hline
$9 \times 11$ &  -36.608(14) & 584.99(48) &  -5349.0(7.4) \\
\hline
$9 \times 12$ &  -38.573(15) & 654.15(54) & -6404.9(8.8) \\
\hline
$9 \times 13$ &  -40.536(17) & 727.01(62) &  -7588(11) \\
\hline
$9 \times 14$ &  -42.493(17) & 803.63(69) & -8908(13) \\
\hline
$9 \times 15$ &  -44.453(19) & 884.06(78) & -10372(15) \\
\hline
$9 \times 16$ &  -46.406(20) & 968.11(88) & -11983(18) \\
\hline
$10 \times 10$ &  -36.629(15) & 585.69(51) & -5358.9(7.9) \\
\hline
$10 \times 11$ &  -38.609(16) & 655.37(57) & -6422.4(9.4) \\
\hline
$10 \times 12$ &  -40.583(17) & 728.79(65) & -7615(11) \\
\hline
$10 \times 13$ &  -42.553(18) & 805.91(73) & -8944(13) \\
\hline
$10 \times 14$ &  -44.523(20) & 886.82(81) & -10418(16) \\
\hline
$10 \times 15$ &  -46.490(21) & 971.52(91) & -12045(18) \\
\hline
$10 \times 16$ &  -48.451(22) & 1059.8(1.0) & -13826(22) \\
\hline
$11 \times 11$ &  -40.595(18) & 729.32(66) & -7624(11) \\
\hline
$11 \times 12$ &  -42.581(18) & 806.97(74) & -8964(13) \\
\hline
$11 \times 13$ &  -44.557(20) & 888.30(84) & -10445(16) \\
\hline
$11 \times 14$ &  -46.530(21) & 973.41(93) & -12080(19) \\
\hline
$11 \times 15$ &  -48.504(22) & 1062.3(1.0) & -13878(22) \\
\hline
$11 \times 16$ &  -50.471(25) & 1154.8(1.2) & -15837(26) \\
\hline
$12 \times 12$ &  -44.570(21) & 888.82(85) & -10455(16) \\
\hline
$12 \times 13$ &  -46.554(23) & 974.34(95) & -12097(19) \\
\hline
$12 \times 14$ &  -48.533(23) & 1063.6(1.0) & -13901(22) \\
\hline
$12 \times 15$ &  -50.512(24) & 1156.7(1.1) & -15875(25) \\
\hline
$12 \times 16$ &  -52.482(26) & 1253.3(1.3) & -18017(30) \\
\hline
$13 \times 13$ &  -48.541(24) &   1064.0(1.1) & -13907(22) \\
\hline
$13 \times 14$ &  -50.526(25) & 1157.4(1.2) & -15886(25) \\
\hline
$13 \times 15$ &  -52.506(26) & 1254.6(1.3) & -18042(29) \\
\hline
$13 \times 16$ &  -54.481(28) & 1355.3(1.4) & -20375(34) \\
\hline
\end{tabular}\\
	\end{center}
\end{table*}

\begin{table*}[h]
	\begin{center}
\begin{tabular}{|c|c|c|c|}
\hline
\, & \, & \, & \\
Loop dimensions  & order $\beta^{-1}$ & order $\beta^{-2}$ & order $\beta^{-3}$ \\
\hline
\hline
$14 \times 14$ &  -52.513(26) & 1254.9(1.3) & -18050(29) \\
\hline
$14 \times 15$ &  -54.496(27) & 1356.2(1.4) & -20397(33) \\
\hline
$14 \times 16$ &  -56.472(29) & 1460.9(1.6) & -22926(39) \\
\hline
$15 \times 15$ &  -56.484(28) & 1461.4(1.5) & -22939(38) \\
\hline
$15 \times 16$ &  -58.459(31) & 1570.1(1.7) & -25672(44) \\
\hline
$16 \times 16$ &  -60.433(33) & 1682.6(1.9) & -28614(51) \\
\hline
\end{tabular}\\
	\end{center}
\end{table*}

\section{Unquenched $n_f=2$ Wilson Loops}

\begin{table*}[h]
	\begin{center}
\begin{tabular}{|c|c|c|c|}
\hline
\, & \, & \, & \\
Loop dimensions  & order $\beta^{-1}$ & order $\beta^{-2}$ & order $\beta^{-3}$ \\
\hline
\hline
$1 \times 1$ &  -2.0001(1) & -1.0883(3) & -2.4124(9) \\
\hline
$1 \times 2$ &  -3.4491(2) & 0.1136(7) & -1.4632(16) \\
\hline
$1 \times 3$ &  -4.8090(4) & 3.1430(15) & 0.0394(28) \\
\hline
$1 \times 4$ &  -6.1509(5) & 7.9558(30) & -0.3321(61) \\
\hline
$1 \times 5$ &  -7.4880(7) & 14.545(5) & -4.949(13) \\
\hline
$1 \times 6$ &  -8.8233(9) &  22.910(8) & -16.163(25) \\
\hline
$1 \times 7$ &  -10.158(1) & 33.047(12) & -36.332(45) \\
\hline
$1 \times 8$ &  -11.492(1) & 44.962(16) & -67.816(73) \\
\hline
$1 \times 9$ &  -12.826(2) & 58.650(21) & -113.00(11) \\
\hline
$1 \times 10$ &  -14.159(2) & 74.118(28) & -174.24(17) \\
\hline
$1 \times 11$ &  -15.493(2) & 91.355(35) & -253.83(24) \\
\hline
$1 \times 12$ &  -16.826(3) & 110.37(4) & -354.26(34) \\
\hline
$1 \times 13$ &  -18.159(3) & 131.16(5) & -477.88(47) \\
\hline
$1 \times 14$ &  -19.492(3) & 153.73(7) & -627.06(64) \\
\hline
$1 \times 15$ &  -20.825(4) & 178.08(8) & -804.16(85) \\
\hline
$1 \times 16$ &  -22.159(4) & 204.21(10) & -1011.6(1.1) \\
\hline
$2 \times 2$ &  -5.4773(5) & 4.7884(23) & 0.6146(51) \\
\hline
$2 \times 3$ &  -7.2455(8) & 12.253(5) & -1.4455(98) \\
\hline
$2 \times 4$ &  -8.9549(10) & 22.494(8) & -12.438(19) \\
\hline
$2 \times 5$ &  -10.647(1) & 35.533(12) & -37.065(37) \\
\hline
$2 \times 6$ &  -12.333(2) & 51.376(18) & -80.018(69) \\
\hline
$2 \times 7$ &  -14.016(2) & 70.022(25) & -145.99(12) \\
\hline
$2 \times 8$ &  -15.697(2) & 91.487(34) & -239.76(19) \\
\hline
$2 \times 9$ &  -17.377(3) & 115.76(5) & -365.94(28) \\
\hline
$2 \times 10$ &  -19.057(3) & 142.84(6) & -529.34(42) \\
\hline
$2 \times 11$ &  -20.736(4) & 172.74(7) & -734.62(59) \\
\hline
$2 \times 12$ &  -22.414(4) & 205.44(9) & -986.34(81) \\
\hline
$2 \times 13$ &  -24.092(5) & 240.95(11) & -1289.4(1.1) \\
\hline
$2 \times 14$ &  -25.771(5) & 279.28(14) & -1648.7(1.5) \\
\hline
$2 \times 15$ &  -27.449(6) & 320.42(16) & -2068.6(1.9) \\
\hline
$2 \times 16$ &  -29.127(7) & 364.38(19) & -2554.1(2.4) \\
\hline
\end{tabular}\\
	\end{center}
\end{table*}

\begin{table*}[p]
	\begin{center}
\begin{tabular}{|c|c|c|c|}
\hline
\, & \, & \, & \\
Loop dimensions  & order $\beta^{-1}$ & order $\beta^{-2}$ & order $\beta^{-3}$ \\
\hline
\hline
$3 \times 3$ &  -9.2206(11) & 24.053(9) & -14.087(26) \\
\hline
$3 \times 4$ &  -11.085(1) & 38.856(14) & -43.287(49) \\
\hline
$3 \times 5$ &  -12.915(2) & 56.814(20) & -94.774(87) \\
\hline
$3 \times 6$ &  -14.732(2) & 77.967(29) & -174.34(14) \\
\hline
$3 \times 7$ &  -16.542(3) & 102.33(4) & -287.74(22) \\
\hline
$3 \times 8$ &  -18.350(3) & 129.94(5) & -440.99(33) \\
\hline
$3 \times 9$ &  -20.154(4) & 160.77(6) & -639.63(47) \\
\hline
$3 \times 10$ &  -21.958(4) & 194.83(8) & -889.56(66) \\
\hline
$3 \times 11$ &  -23.760(5) & 232.12(10) & -1196.7(9) \\
\hline
$3 \times 12$ &  -25.562(5) & 272.63(12) & -1566.5(1.2) \\
\hline
$3 \times 13$ &  -27.363(7) & 316.37(15) & -2005.1(1.6) \\
\hline
$3 \times 14$ &  -29.164(7) & 363.37(18) & -2518.7(2.1) \\
\hline
$3 \times 15$ &  -30.966(7) & 413.60(21) & -3112.4(2.7) \\
\hline
$3 \times 16$ &  -32.765(8) & 467.06(25) & -3792.2(3.4) \\
\hline
$4 \times 4$ &  -13.046(2) & 58.048(21) & -98.171(93) \\
\hline
$4 \times 5$ &  -14.954(2) & 80.475(30) & -183.47(15) \\
\hline
$4 \times 6$ &  -16.840(3) & 106.24(4) & -305.51(24) \\
\hline
$4 \times 7$ &  -18.715(4) & 135.40(5) & -470.56(36) \\
\hline
$4 \times 8$ &  -20.585(5) &    168.00(7) & -685.39(52) \\
\hline
$4 \times 9$ &  -22.452(5) & 204.01(8) & -956.02(71) \\
\hline
$4 \times 10$ &  -24.315(5) & 243.45(10) & -1288.9(9) \\
\hline
$4 \times 11$ &  -26.177(6) & 286.33(12) & -1690.7(1.2) \\
\hline
$4 \times 12$ &  -28.038(7) & 332.66(14) & -2167.6(1.6) \\
\hline
$4 \times 13$ &  -29.898(7) & 382.40(17) & -2725.9(2.1) \\
\hline
$4 \times 14$ &  -31.758(7) & 435.63(20) & -3372.5(2.7) \\
\hline
$4 \times 15$ &  -33.617(8) & 492.31(24) & -4113.5(3.4) \\
\hline
$4 \times 16$ &  -35.477(8) & 552.42(28) & -4955.1(4.2) \\
\hline
$5 \times 5$ &  -16.918(3) & 107.28(4) & -310.20(27) \\
\hline
$5 \times 6$ &  -18.851(3) & 137.46(6) & -481.94(40) \\
\hline
$5 \times 7$ &  -20.768(4) & 171.11(7) & -705.30(56) \\
\hline
$5 \times 8$ &  -22.678(4) & 208.30(9) & -987.54(80) \\
\hline
$5 \times 9$ &  -24.582(6) & 248.99(11) & -1334.7(1.1) \\
\hline
$5 \times 10$ &  -26.482(6) & 293.22(14) & -1753.9(1.4) \\
\hline
$5 \times 11$ &  -28.379(7) & 341.01(16) & -2252.0(1.8) \\
\hline
$5 \times 12$ &  -30.276(8) & 392.36(19) & -2835.7(2.4) \\
\hline
$5 \times 13$ &  -32.170(8) & 447.25(23) & -3511.6(3.1) \\
\hline
$5 \times 14$ &  -34.065(9) & 505.76(26) & -4287.1(3.8) \\
\hline
$5 \times 15$ &  -35.959(10) & 567.83(31) & -5168.9(4.7) \\
\hline
$5 \times 16$ &  -37.853(12) & 633.47(35) & -6162.9(5.8) \\
\hline
$6 \times 6$ &  -20.820(5) & 172.00(8) & -711.20(64) \\
\hline
$6 \times 7$ &  -22.769(5) & 210.03(10) & -1000.2(9) \\
\hline
$6 \times 8$ &  -24.709(6) & 251.64(12) & -1356.7(1.2) \\
\hline
$6 \times 9$ &  -26.639(6) & 296.80(14) & -1787.0(1.5) \\
\hline
$6 \times 10$ &  -28.565(6) & 345.57(17) & -2298.4(1.9) \\
\hline
\end{tabular}\\
	\end{center}
\end{table*}

\begin{table*}[p]
	\begin{center}
\begin{tabular}{|c|c|c|c|}
\hline
\, & \, & \, & \\
Loop dimensions  & order $\beta^{-1}$ & order $\beta^{-2}$ & order $\beta^{-3}$ \\
\hline
\hline
$6 \times 11$ &  -30.488(7) & 397.96(20) & -2897.6(2.5) \\
\hline
$6 \times 12$ &  -32.408(8) & 454.00(23) & -3592.4(3.1) \\
\hline
$6 \times 13$ &  -34.326(9) & 513.66(27) & -4389.8(4.0) \\
\hline
$6 \times 14$ &  -36.243(10) & 577.00(31) & -5296.2(4.8) \\
\hline
$6 \times 15$ &  -38.161(11) & 643.97(36) & -6318.2(5.8) \\
\hline
$6 \times 16$ &  -40.077(12) & 714.60(40) & -7464.1(7.0) \\
\hline
$7 \times 7$ &  -24.746(5) & 252.41(12) & -1363.2(1.2) \\
\hline
$7 \times 8$ &  -26.707(6) & 298.37(15) & -1801.5(1.6) \\
\hline
$7 \times 9$ &  -28.658(7) & 347.89(18) & -2321.9(2.1) \\
\hline
$7 \times 10$ &  -30.604(7) & 401.08(21) & -2932.2(2.7) \\
\hline
$7 \times 11$ &  -32.545(8) & 457.91(24) & -3639.4(3.3) \\
\hline
$7 \times 12$ &  -34.481(9) & 518.42(27) & -4450.8(4.1) \\
\hline
$7 \times 13$ &  -36.419(10) & 582.64(32) & -5374.8(5.0) \\
\hline
$7 \times 14$ &  -38.353(10) & 650.58(36) & -6417.0(6.0) \\
\hline
$7 \times 15$ &  -40.288(11) & 722.20(41) & -7584.6(7.3) \\
\hline
$7 \times 16$ &  -42.219(11) & 797.55(46) & -8886.7(8.6) \\
\hline
$8 \times 8$ &  -28.689(7) & 348.65(19) & -2329.4(2.2) \\
\hline
$8 \times 9$ &  -30.658(8) & 402.50(22) & -2948.0(2.7) \\
\hline
$8 \times 10$ &  -32.619(9) & 460.03(26) & -3665.1(3.5) \\
\hline
$8 \times 11$ &  -34.574(9) & 521.22(29) & -4487.5(4.2) \\
\hline
$8 \times 12$ &  -36.526(10) & 586.12(33) & -5422.8(5.1) \\
\hline
$8 \times 13$ &  -38.476(11) & 654.75(38) & -6479.3(6.2) \\
\hline
$8 \times 14$ &  -40.424(11) & 727.14(43) & -7662.9(7.5) \\
\hline
$8 \times 15$ &  -42.370(12) & 803.25(49) & -8981.2(9.0) \\
\hline
$8 \times 16$ &  -44.314(13) & 883.16(55) & -10444(11) \\
\hline
$9 \times 9$ &  -32.639(9) & 460.64(27) & -3672.9(3.5) \\
\hline
$9 \times 10$ &  -34.615(10) & 522.46(31) & -4504.0(4.5) \\
\hline
$9 \times 11$ &  -36.582(11) & 587.94(34) & -5448.6(5.4) \\
\hline
$9 \times 12$ &  -38.546(12) & 657.15(39) & -6514.7(6.4) \\
\hline
$9 \times 13$ &  -40.507(13) & 730.09(45) & -7710.5(7.9) \\
\hline
$9 \times 14$ &  -42.464(13) & 806.83(50) & -9042.2(9.2) \\
\hline
$9 \times 15$ &  -44.420(13) & 887.30(57) & -10517(11) \\
\hline
$9 \times 16$ &  -46.373(14) & 971.61(63) & -12147(13) \\
\hline
$10 \times 10$ &  -36.602(11) & 588.54(36) & -5457.1(5.6) \\
\hline
$10 \times 11$ &  -38.581(12) & 658.28(40) & -6532.0(6.5) \\
\hline
$10 \times 12$ &  -40.551(12) & 731.74(45) & -7736.1(7.7) \\
\hline
$10 \times 13$ &  -42.520(13) & 808.92(51) & -9077.9(9.4) \\
\hline
$10 \times 14$ &  -44.488(14) & 889.93(57) & -10565(11) \\
\hline
$10 \times 15$ &  -46.453(15) & 974.70(64) & -12204(13) \\
\hline
$10 \times 16$ &  -48.414(15) & 1063.31(71) & -14005(15) \\
\hline
$11 \times 11$ &  -40.566(12) & 732.25(44) & -7744.1(7.8) \\
\hline
$11 \times 12$ &  -42.549(13) & 809.94(49) & -9094.4(9.0) \\
\hline
$11 \times 13$ &  -44.522(14) & 891.34(56) & -10590(11) \\
\hline
$11 \times 14$ &  -46.496(14) & 976.61(61) & -12240(12) \\
\hline
\end{tabular}\\
	\end{center}
\end{table*}

\begin{table*}[h]
	\begin{center}
\begin{tabular}{|c|c|c|c|}
\hline
\, & \, & \, & \\
Loop dimensions  & order $\beta^{-1}$ & order $\beta^{-2}$ & order $\beta^{-3}$ \\
\hline
\hline
$11 \times 15$ &  -48.465(15) & 1065.59(69) & -14049(15) \\
\hline
$11 \times 16$ &  -50.433(16) & 1158.43(75) & -16030(17) \\
\hline
$12 \times 12$ &  -44.535(13) & 891.87(55) & -10600(11) \\
\hline
$12 \times 13$ &  -46.515(14) & 977.47(63) & -12257(13) \\
\hline
$12 \times 14$ &  -48.495(15) & 1066.92(68) & -14076(15) \\
\hline
$12 \times 15$ &  -50.471(16) & 1160.1(8) & -16063(17) \\
\hline
$12 \times 16$ &  -52.444(16) & 1257.1(8) & -18232(20) \\
\hline
$13 \times 13$ &  -48.501(15) & 1067.22(72) & -14084(16) \\
\hline
$13 \times 14$ &  -50.486(16) & 1160.84(78) & -16080(18) \\
\hline
$13 \times 15$ &  -52.465(17) & 1258.13(88) & -18252(21) \\
\hline
$13 \times 16$ &  -54.441(18) & 1359.29(95) & -20614(23) \\
\hline
$14 \times 14$ &  -52.477(17) & 1258.6(9) & -18261(20) \\
\hline
$14 \times 15$ &  -54.458(18) & 1360.02(96) & -20627(24) \\
\hline
$14 \times 16$ &  -56.438(19) & 1465.3(1.0) & -23190(27) \\
\hline
$15 \times 15$ &  -56.444(20) & 1465.5(1.1) & -23191(28) \\
\hline
$15 \times 16$ &  -58.426(21) & 1574.8(1.2) & -25960(31) \\
\hline
$16 \times 16$ &  -60.408(22) & 1688.2(1.3) & -28944(35) \\
\hline
\end{tabular}\\
	\end{center}
\end{table*}

\end{appendix}

\end{document}